\documentclass{aa}
\usepackage{graphicx}
\usepackage[varg]{txfonts}

\usepackage{natbib}
\bibpunct{(}{)}{;}{a}{}{,}
\usepackage{subcaption}
\usepackage{tablefootnote}
\usepackage{threeparttable}
\usepackage{float}
\usepackage{url}
\makeatletter
\renewcommand*\aa@pageof{, page \thepage{} of \pageref*{LastPage}}
\makeatother

\newcommand \Cp[1]{[\ion{\element[][#1]{C}}{II}]}

\newcommand \OI[1]{[\ion{O}{I}] #1~$\mu$m}
\newcommand \Oone {[\ion{O}{I}]}
\newcommand \um {$\mu$m}

\usepackage{hyperref}
\hypersetup{
    colorlinks=true,
    linkcolor=blue,
    citecolor=blue,
    urlcolor = blue,
    unicode=true
}
\usepackage{amsmath}

\title{ The [\ion{O}{I}] fine structure line profiles in Mon~R2 and M17~SW:  
The puzzling nature of cold foreground material identified by \Cp{12} self-absorption}
\author{C. Guevara \inst{1,2} \and J. Stutzki \inst{1}  \and V. Ossenkopf-Okada \inst{1} \and U. Graf \inst{1} \and Y. Okada \inst{1} \and N. Schneider \inst{1} \and P.F. Goldsmith \inst{3} \and J.P. P\'erez-Beaupuits \inst{4,6,7} \and S. Kabanovic \inst{1} \and M. Mertens \inst{1} \and N. Rothbart\inst{5} \and  R. G\"{u}sten \inst{6}  } 

\institute{I. Physikalisches Institut, Universit\"{a}t zu K\"{o}ln, Z\"{u}lpicher Str. 77, 50937 K\"{o}ln, Germany \\
              \email{guevara@ph1.uni-koeln.de} \and
              Instituto de Astronom{\'i}a, Universidad Cat{\'o}lica del Norte, Av. Angamos 0610, 1270398 Antofagasta, Chile
              \and
              Jet Propulsion Laboratory, California Institute of Technology, 4800 Oak Grove Drive, Pasadena, CA 91109-8099, USA                   \and
European Southern Observatory, Alonso de C\'{o}rdova 3107, Vitacura, Santiago, Chile              \and
Institute of Optical Sensor Systems, German Aerospace Center (DLR), Rutherfordstr. 2, 12489 Berlin, Germany              
\and              
             Max-Planck-Institut f\"{u}r Radioastronomie, Auf dem H\"{u}gel 69, D-53121 Bonn, Germany
\and
Centro de Astro-Ingenier{\'i}a UC, Instituto de Astrof{\'i}sica, Pontificia Universidad Cat{\'o}lica de Chile, Avda Vicu\~na Mackenna 4860, Macul, Santiago, Chile             
             }

\date{Received date /
Accepted date }

\begin{document}

 \abstract
{Recent studies of the optical depth comparing \Cp{12} and \Cp{13} line profiles in Galactic star-forming regions have revealed strong self-absorption in \Cp{12} by low excitation foreground material. This implies a high column density for C$^+$, corresponding to equivalent A$_{\mathrm{V}}$ values of a few (up to about 10) mag.
 }
{
As the nature and origin of such a great column of cold C$^+$ foreground gas are difficult to determine, it is essential to constrain the physical conditions of this material.
}
{
We conducted high-resolution observations of \OI{63} and \OI{145} lines in M17~SW and Mon~R2. The \OI{145} transition traces warm PDR-material, while the \OI{63} line traces the foreground material, as manifested by the absorption dips. 
}
{A comparison of both \Oone~line profiles with \Cp{} isotopic lines confirm warm PDR-origin background emission and a significant column of cold foreground material, causing the self-absorption to be visible in the \Cp{12} and \OI{63} profiles. In M17~SW, the \element[+][]{C} and \element[0][]{O} column densities are comparable for both layers. Mon~R2 exhibits larger \element[0][]{O} columns compared to \element[+][]{C}, indicating additional material where the carbon is neutral or in molecular form. Small-scale spatial variations in the foreground absorption profiles and the large column density ($\sim$ 10$^{18}$ cm$^{-2}$) of the foreground material suggest the emission is coming from high-density regions associated with the cloud complex -- and not a uniform diffuse foreground cloud. 
}
{The analysis confirms that the previously detected intense \Cp{} foreground absorption is attributable to a large column of low-excitation dense atomic material, where carbon is ionized and oxygen is in a neutral atomic form.
}

\titlerunning{\ion{O}{I}] fine structure line profiles in Mon~R2 and M17~SW}
\maketitle 

\section{Introduction} \label{Introduction}

The fine structure emission lines of [\ion{O}{I}], together with the \Cp{} 158 $\mu$m emission line and high-J CO lines, are the main cooling lines for photodissociation regions \citep[PDRs,][]{1985ApJ...291..722T,1999RvMP...71..173H} in the interstellar medium (ISM). The spin-orbit coupling in neutral atomic oxygen leads to three fine-structure levels and, correspondingly, two \Oone~fine structure transitions. The lower transition, \Oone~$^3$P$_1\rightarrow^3$P$_2$, has a wavelength of 63.2~$\mu$m, corresponding to an energy of its upper level of 227.7~K; the collisional rate coefficients with \element[][]{H} give a critical density at 77~K of 7.8$\times$10$^5$~cm$^{-3}$, and with \element[][]{H}$_2$ as collision partner, the critical density is 5$\times$10$^5$~cm$^{-3}$ \citep{2019ApJ...887...54G}. The upper transition, \Oone~$^3$P$_0\rightarrow^3$P$_1$, has a wavelength of 145.5~$\mu$m. Its upper state energy is 326.6~K above the ground state, or 98.9~K above the mid-level, with a critical density of 5.8$\times$10$^6$~cm$^{-3}$ for \element[][]{H}$_2$ as collision partner \citep{2019ApJ...887...54G}. The \Oone~fine structure emission from PDRs is thus bright if the gas is dense. It is, therefore, together with the \Cp{} 158~$\mu$m fine structure line, commonly used for tracing the warm and dense gas in star-forming regions, locally and out to high redshift galaxies. The \OI{63} transition rapidly reaches a high optical depth in low-temperature gas (T$\ll$ 230~K), so high that the intensity is no longer a measure of the column density. The fine structure lines of the oxygen isotopes (in particular {\ion{\element[][18]{O}}{}}) are so close in frequency that they blend with the Doppler-broadened main isotope transition in astronomical sources so that the isotopic line ratios cannot be used to determine the \Oone~ optical depth. Early indications of high optical depth have been reported by \citet{1983ApJ...265L...7S} and \citet{1996ApJ...464L..83B} based on a comparison of the integrated line intensity ratio of the \Oone~lines, but the optical depth could be only inferred indirectly. High signal-to-noise (S/N) line profiles are required to directly measure the fine structure line optical depth, showing saturation or self-absorption. With the high spectral resolution and high S/N available with the GREAT \citep{2012A&A...542L...1H} instrument on the Stratospheric Observatory for Infrared Astronomy \citep[SOFIA,][]{2012ApJ...749L..17Y,2018JAI.....740011T}, several authors \citet[]{leurini15, 2018A&A...617A..45S,2019A&A...626A.131M,2020MNRAS.497.2651K,2021ApJ...916....6G} have recently reported velocity resolved \OI{63} observations displaying complex line profiles. A recent survey of \OI{63} line profiles towards 12 star-forming cores in the Milky Way by \citet{2021ApJ...916....6G} showed self-absorption by foreground material for half of the sources. In addition, high optical depth and self-absorption in \Oone~ are often invoked to explain the observed too-low \OI{63} intensities compared to model predictions from other observed lines.\par

In a recent study of several Galactic star-forming regions comparing the \Cp{12} and \Cp{13} line profiles at high S/N \citep{2020A&A...636A..16G}, we have shown that the \Cp{} 158~$\mu$m emission is optically thick in a wide range of physical conditions. The two sources, Monoceros~R2 (Mon~R2) and M17~SW, were studied and showed deep and narrow self-absorption in the line profiles against the bright and broad background line emission. The analysis of those spectra showed that the bright background line emission, derived from the optically thin \Cp{13} line profile and compatible with originating in PDRs, is absorbed by a large column density of ionized carbon in cold foreground material, with a low excitation temperature (T$_{\mathrm{ex}} \lesssim$ 25~K). A lower limit for the foreground material, derived under the assumption that all carbon is in the form of \element[+][]{C}, gives a corresponding visual extinction of several (up to ten) magnitudes. The amount of material can be considerably larger if the carbon is only partially ionized and the foreground gas contains more carbon in the form of neutral atoms or bound in molecules, particularly CO. However, the velocities and line widths of these foreground absorption features do not match with any features in the spectral lines of carbon monoxide, indicating that the foreground absorbing material presents a component of the ISM that is separate from the molecular gas traced by CO. To present knowledge, ionized carbon is present mainly in PDR layers at temperatures above 60. Hence, the nature of such large amounts of ionized carbon in low excitation layers of gas is very puzzling. Low excitation may imply low density (well below the critical density of \Cp{}), but such diffuse gas would need to fill large volumes to explain the observed total column density. \citet{2022A&A...659A..36K} presented a scenario for the source RCW~120, where the carbon is present in a diffuse envelope with neutral hydrogen.\par

Mon~R2 is a star-forming region at a distance of 778~pc \citep{2019ApJ...879..125Z}. The region contains a reflection nebula, and the UC\ion{H}{II} region is surrounded by several PDRs with different physical conditions. The molecular cloud associated with Mon~R2 has a hub-filament structure and contains clumps of density up to 10$^6$ cm$^{-3}$. The radiation field in the interface region between the \ion{H}{II} region and the cloud is about 10$^{5}$ G$_{\mathrm{o}}$ \citep{2012A&A...544A.110P}. The sources have been studied through several atomic and molecular tracers \citep[e.g.,][]{2012A&A...544A.110P,2013A&A...554A..87P,2014A&A...569A..19T,2017A&A...607A..22R,2019A&A...629A..81T}. \par

M17 is one of the brightest and most massive star-forming regions in the Galaxy, located at a distance of 1.9~kpc \citep{2019ApJ...874...94W}. M17~SW is the sub-region located in the southwest, where an \ion{H}{II} region is localized and is associated with a giant molecular cloud and PDR interface. M17~SW is considered a prototype of an edge-on PDR. The \ion{H}{II} region is ionized by a highly obscured (with a visual extinction A$_{\mathrm{V}}$ > 10~mag) cluster of many (>100) OB stars \citep{2008ApJ...686..310H}. The gas near the \ion{H}{II} region is distributed in high density clumps ($n$ $\leq$ 10$^5$~cm$^{-3}$ ) embedded in interclump material ($n$ $\sim$ 10$^3$~cm$^{-3}$) surrounded by diffuse gas ($n$ $\sim$ 300 cm$^{-3}$), irradiated by a strong UV field of G$_{\mathrm{o}}$ = 5.6$\times$10$^{4}$ \citep{1992ApJ...390..499M}. Recently, using non-velocity resolved observations of several infrared lines such as \Cp{}, both \Oone~lines, [\ion{\element{O}}{III}], or [\ion{\element{N}}{III}] from FIFI-LS \citep{2018JAI.....740004C,2018JAI.....740003F},  \citet{2023ApJ...945...29K} derived hydrogen nuclei density and UV radiation maps, with an average hydrogen density of 10$^{5.9}$ cm$^{-3}$ for the molecular cloud. They found that the ionization and photodissociation fronts are nearly merged with a sharp density jump from the ionized region to the neutral one. \par

To characterize the nature of the absorbing layer in these two sources, Mon~R2 and M17~SW, we followed up on the investigation we started in \citet{2020A&A...636A..16G} by observing the two sources in both \Oone~fine structure lines with the upGREAT\footnote{upGREAT is a development by the MPI für Radioastronomie and KOSMA/Universität zu Köln, in cooperation with the DLR Institut für Optische Sensorsysteme.} instrument \citep{2016A&A...595A..34R} on board SOFIA along the same lines of sight, as previously observed in \Cp{}. \par 

The paper is organized as follows. Section~\ref{Observations} details the observational setup and data reduction. Section~\ref{Results} shows the observed \OI{145} and \OI{63} line profiles and a comparison against \Cp{12} and \Cp{13} profiles. Section~\ref{multicomponent} presents a multi-component double-layered Gaussian profile analysis using both \Oone~lines to estimate oxygen column densities and excitation temperature. Section~\ref{Discussion} discusses the results and the possible nature of the background and foreground layers. Finally, Sect.~\ref{Summary} summarizes the present work. 

\section{Observations and data reduction} \label{Observations}

\begin{figure}
   \centering
   \includegraphics[width=1\hsize]{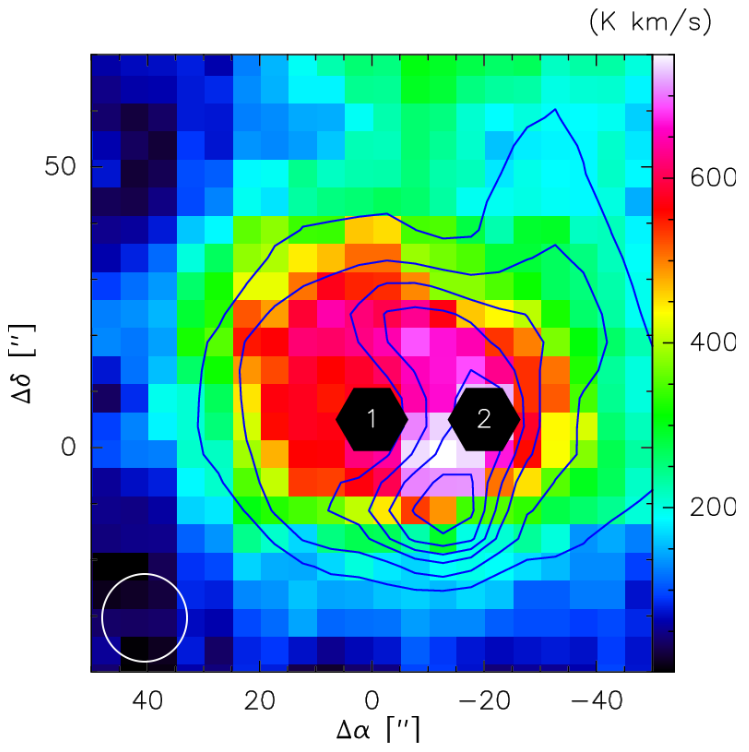}
      \caption{Mon~R2 \Cp{} integrated intensity map for the velocity range between 0 and 30~km s$^{-1}$  with the two \Cp{} single pointings (in black) from \citet{2020A&A...636A..16G}. \OI{63} integrated intensity map in blue contours covering the same velocity range of \Cp{} (levels at 100, 150, 200, 240, 270, and 300~K km s$^{-1}$). The white circle represents the FWHM beam size of both maps, 15\arcsec.
              }
         \label{fig:monr2pos}
   \end{figure}

We observed the \OI{63} and \OI{145} fine-structure lines in Mon~R2 and M17~SW with the 7-pixel High-Frequency Array (HFA) and the 7x2-pixel Low-Frequency Array (LFA) arrays of the upGREAT receiver on board SOFIA. Mon~R2 was observed in December 2018, and M17~SW in June 2019 and April 2022. For Mon~R2, the LFA was tuned to observe the \OI{145} in the H-polarization array (LFAH) and \Cp{} 158~$\mu$m line in the V-polarization (LFAV) simultaneously. The single-sideband (SSB) system temperatures at the source velocity (T$_{\mathrm{sys}}$) were 3500 and 3050~K, respectively. In parallel, the HFA was tuned to \OI{63}, with a T$_{\mathrm{sys}}$(SSB) of 4700~K. The parameters are summarized in Table~\ref{table:tsys}. As the former \Cp{} spectra \citep{2020A&A...636A..16G} were observed with the old single-pixel configuration of the GREAT instrument at two discrete positions, we had no information on the spatial variation of the \Cp{} line profiles. Hence, we first observed a small, fully sampled map (for the three lines) of 180\arcsec$\times$180\arcsec~extent in on-the-fly mode with a grid separation of 3\arcsec~in the horizontal and vertical directions. Then, we performed deep integrations in total power mode of 5~min ON-source time for the two positions previously observed in \Cp{} (Fig.~\ref{fig:monr2pos}). Appendix~\ref{app:dpeac} lists the coordinates of each position in detail and the OFF positions used. The two \Cp{} positions were selected to follow the peak emission. The average precipitable water vapor column was 7~\um. The OFF position presented some weak contamination of $\sim$2~K, corrected through the same procedure for \Cp{} as described by \citet[][]{2020A&A...636A..16G}, Appendix B, fitting a Gaussian profile to the OFF position spectra and then adding the fitted profile into the ON spectra to recover the lost emission.  \par

\begin{table*}[ht]
  \centering
    \caption{Observational parameters.}
    \label{table:tsys}
  \begin{threeparttable}  
  \begin{tabular}{l | l l c c c c c c c }
      \hline
      \hline
 Sources  & RA & DEC & LFAH  & T$_{\mathrm{sys}}$  & LFAV  & T$_{\mathrm{sys}}$ & HFA & T$_{\mathrm{sys}}$ & <pwv>\tnote{a}\\
          & (h:m:s) & (\degr:$'$:$''$) &    & (K) & & (K) & & (K) & (\um) \\
            \hline
Mon~R2  & 06:07:46.2 & $-$06:23:08.0 & \OI{145} & 3700 & \Cp{} 158 \um & 3150 & \OI{63} & 4800 & 7  \\
M17~SW$_{\mathrm{June 2019}}$  & 18:20:27.6 & $-$16:12:00.9 & \OI{145} & 3000 & \Cp{} 158 \um & 2100 & - & - & 8  \\
M17~SW$_{\mathrm{April 2022}}$  & 18:20:27.6 & $-$16:12:00.9 & - & - & - & - & \OI{63} & 5200 & 6.5    \\
        \hline
\end{tabular}
\begin{tablenotes}\footnotesize
\item[a] Average over all pixels of the precipitable water vapor column.
\end{tablenotes}
\end{threeparttable}
 \end{table*}

\begin{figure}
   \centering
   \includegraphics[width=1\hsize]{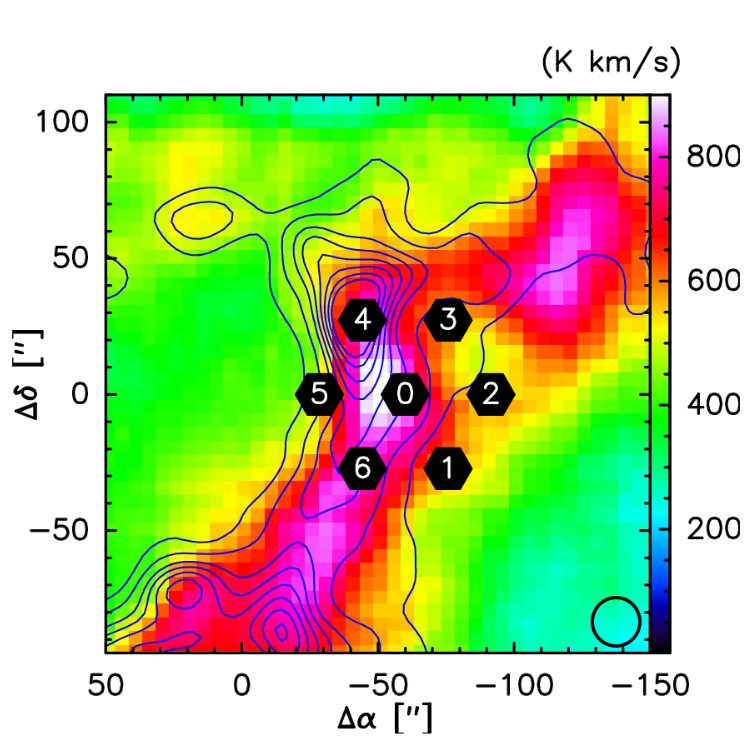}
      \caption{M17~SW \Cp{} integrated intensity map for the velocity range between 0 and 40~km s$^{-1}$ with the original \Cp{} single pointings (in black) from \citet{2020A&A...636A..16G} (equivalent to a LFA array pointing). \OI{63} integrated intensity map in blue contours covering the same velocity range of \Cp{} (levels at 40, 70, 100, 120, 140, and 160 and 200~K km s$^{-1}$). The black circle shows the FWHM beam size for both maps, 15\arcsec. 
              }
         \label{fig:M17array}
   \end{figure}

In June 2019, we observed a fully sampled small \Oone~map for M17~SW using upGREAT in OTF mode with a grid separation of 3\arcsec~in horizontal and vertical directions. The LFA was tuned to simultaneously observe the \OI{145} in LFAH and \Cp{} 158~$\mu$m in LFAV, with a T$_{\mathrm{sys}}$(SSB) at the velocity of the source of 3000 and 2000~K, respectively. The map has an extent of 220\arcsec$\times$220\arcsec. The average precipitable water vapor was 8~\um. The HFA was tuned to \OI{63}, but the telluric line was located in the central velocities of the emission profile, rendering the observations useless. Hence, we repeated the same observations in April 2022 at a different time to avoid the telluric line contamination for the \OI{63} emission. The observations were successful, but this time \OI{145} was the line affected by the telluric feature in the central velocities. The average precipitable water vapor was 6.5~\um with a T$_{\mathrm{sys}}$(SSB) at 63~\um~of 5100~K. The \Cp{} integrated intensity map shown in Fig.~\ref{fig:M17array} is the combination of the observations referred above and a map observed within the SOFIA Feedback Legacy Project\footnote{\url{https://feedback.astro.umd.edu/}} \citep[][]{2020PASP..132j4301S}. We note that the full \Cp{} map will be discussed in a separate publication\footnote{For details on the observing scheme, see \citet{{2020PASP..132j4301S}}.}. The positions for the analysis below were selected following the \Cp{} peak emission along the ridge. The seven positions were previously observed in \Cp{12} and \Cp{13}, already analyzed in \citet{2020A&A...636A..16G}. The OFF position was apparently free of emission. \par

The data were calibrated to an intensity scale in main beam brightness temperature, T${_\mathrm{mb}}$, with the {\it kalibrate} task \citep{2012A&A...542L...4G}, part of the standard GREAT pipeline. Then, we subtracted baselines with the CLASS 90 package, part of the GILDAS\footnote{\url{https://www.iram.fr/IRAMFR/GILDAS/}} software, and resampled the data to 0.3~km s$^{-1}$ channel width, the same as used for the \Cp{} data.\par 

For Mon~R2, the atmospheric 63 $\mu$m atomic oxygen line at the time of the observation was located at -10~km s$^{-1}$ LSR-velocity, sufficiently far away from the emission of the source; hence, it did not affect the analysis of the line profile. For an inter-comparison, we have convolved the three maps to a 15~\arcsec~beam size from the nominal 6.3\arcsec~for the \OI{63} and 14.1\arcsec~for the \OI{145} and \Cp{} lines. For M17~SW, the maps observed at different times of the year allowed us to avoid the telluric line. Hence, both \Oone~maps were convolved to a 15~\arcsec~beam size on a 5\arcsec~grid with a velocity resolution of 0.3~km s$^{-1}$. All the maps were gridded and convolved into a joint resolution to compare them in identical conditions, avoiding the difference in intensities given by the dissimilar beam sizes. 

\section{\Oone~line profiles} \label{Results}

\begin{figure}
   \centering
   \includegraphics[width=0.95\hsize]{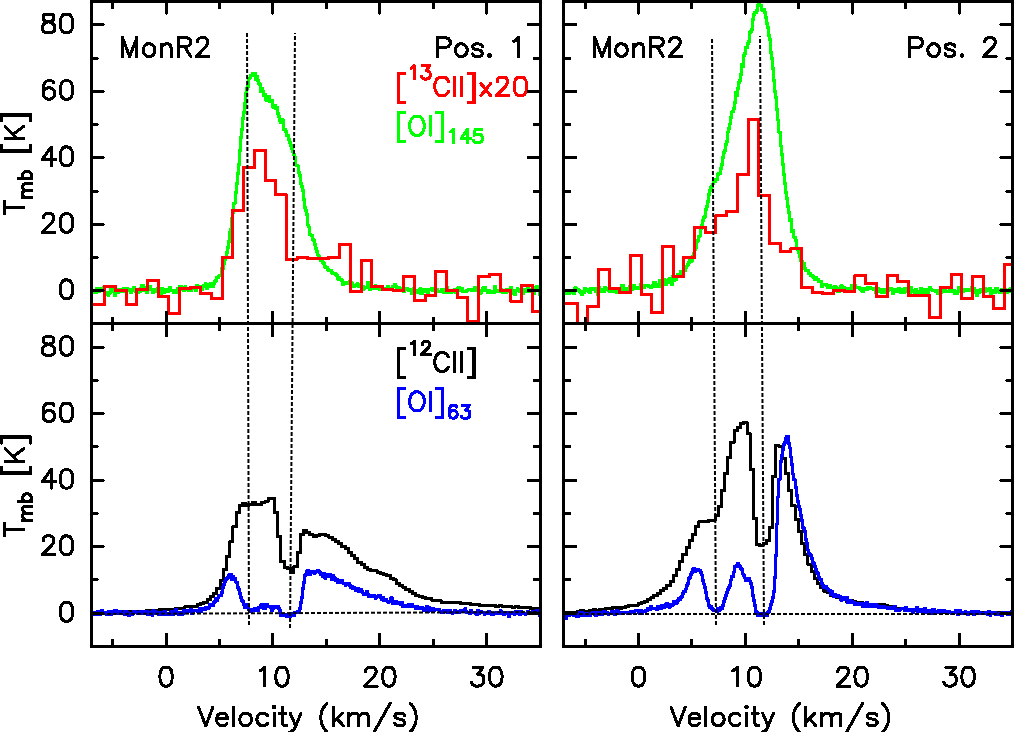}
      \caption{Mon~R2 \OI{63}, \OI{145}, \Cp{12} and \Cp{13} spectra for the single pointings previously observed in \Cp{}. All spectra have been convolved to a beam size of 15\arcsec. The vertical dashed lines at the 8 and 11~km s$^{-1}$ absorption dip are present in both spectra, below the horizontal dashed line at 0~K. The \Cp{13} has been smoothed in velocity for a matter of display.
              }
         \label{fig:MONR2OICII}
   \end{figure}

The \Cp{} and \Oone~emission in Mon~R2 is compact (see Fig.~\ref{fig:monr2pos}), extending over $\approx$ 60\arcsec~in all three lines, in contrast to M17~SW, where the emission in both lines is extended along the PDR edge (see Fig.~\ref{fig:M17array}). Figure~\ref{fig:MONR2OICII} compares the line profiles of the newly observed \Oone~spectra in Mon~R2 obtained in deep integration and the previously observed \Cp{12} and \Cp{13} spectra, the \Cp{13} profiles shown are the average of the two outer hfs-satellites, as explained in \citet{2020A&A...636A..16G}. The \OI{145} profiles are very similar to those of the \Cp{13}~line, though the latter, being very weak, have a lower S/N. The line profiles agree in peak position and width but also show similarities in the detailed profile structure, which is composed of two Gaussian components overlapping in velocity and added together, as expected for optically thin emission. In contrast, the \OI{63} profile shows deep absorption notches at 8 and 12~km s$^{-1}$ velocity and possibly a number of additional weaker ones. The overlay with the \Cp{12} profiles shows that the \Oone~absorption matches in center velocity and line width with the self-absorption notches visible in the \Cp{12} line.\par
   
\begin{figure*}
   \centering
   \includegraphics[width=0.85\hsize]{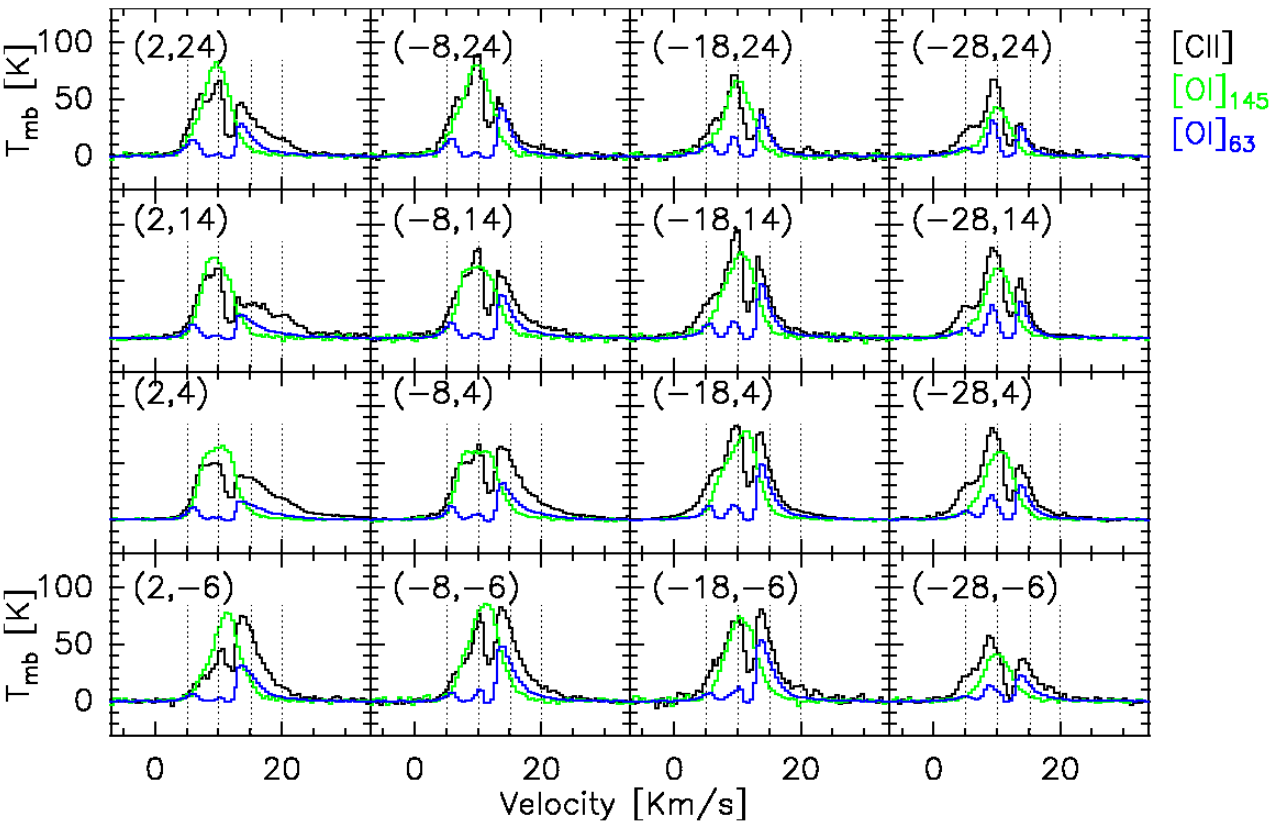}
      \caption{Mon~R2 \OI{63}, \OI{145}, and \Cp{} mosaic spectra for the central area of the source. The dotted lines are located at 5, 10, 15, and 20~km s$^{-1}$, respectively. The angular resolution is the same for \Cp{}, \OI{145} and \OI{63} (15\arcsec), with a grid spacing of 10\arcsec. The offset coordinates of each spectrum with respect to the source position are given in the top-left corner of each box.}      
         \label{fig:MONR2mosaic}
   \end{figure*}

The match in the absorption notches between \Cp{12} and \OI{63} is present not only in the two positions mentioned above. As shown in Fig.~\ref{fig:MONR2mosaic}, the self-absorption notches are strong across the source and show spatial variation over distances as close as 10\arcsec~to 20\arcsec, corresponding to 0.6 to 0.12~pc at the distance of Mon~R2, with changes in intensity at this distances (see Sect.~\ref{foreground_physical} for a discussion of this variation).\par

The absorption to negative intensities (after baseline subtraction) at 11~km s$^{-1}$ can be explained by a weak 63 \um~dust continuum. Additional diffuse gas is present in the line of sight. \citet{2013A&A...554A..87P} found foreground absorption through observations of small hydrocarbons at the referenced velocity, with an equivalent A$_{\mathrm{V}}$ of 1~magnitude, detached from the source. The absorption by this component is insufficient to explain the absorption dips. However, it may weakly contribute to the absorption at this velocity (see Sect.~\ref{Results} for an analysis of the absorption dips). The self-absorption notches strongly reduce the \OI{63} integrated intensity. The \OI{63} emission is almost completely absorbed out between 10 and 12~km s$^{-1}$. \par
 
\begin{figure}
   \centering
   \includegraphics[width=0.95\hsize]{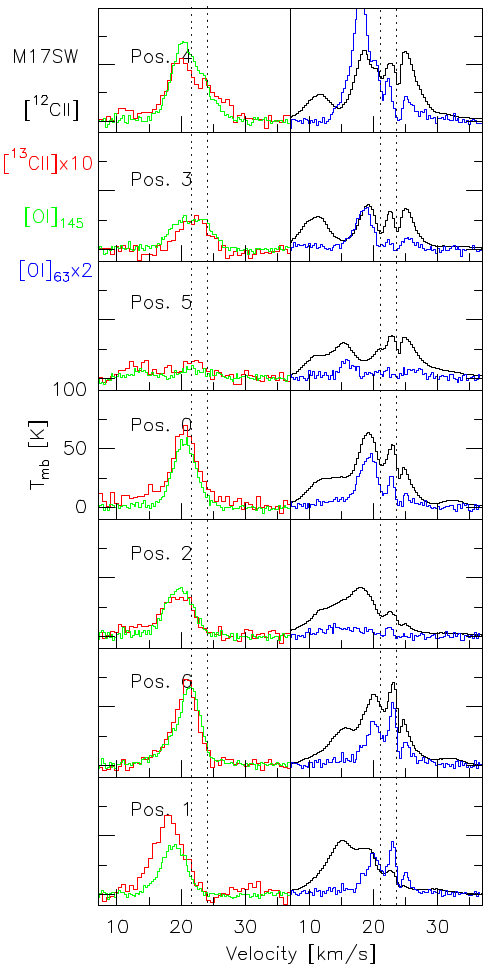}
      \caption{Same as Fig.~\ref{fig:MONR2OICII}, but for M17~SW. The dotted lines mark the two absorption dips present for both fine structure lines at 21 and 24~km s$^{-1}$. 
              }
         \label{fig:M17OICII}
   \end{figure}

The comparison of the line profiles shows a similar behavior for M17~SW (Fig.~\ref{fig:M17OICII}): the \OI{145} line profile is very similar to the formerly observed \Cp{13} profiles (where, in this case, the similarity can be confirmed in detail, because of the high S/N of the deep \Cp{13} integrations). The \OI{63} profiles are heavily affected by self-absorption, and the absorption features are well correlated in position and width with those identified in \Cp{12} (see the next section for more details). Only the lower velocity component at around 11~km s$^{-1}$ in the \Cp{12} line is not reproduced in \OI{63}. As discussed in \citet{2020A&A...636A..16G}, this component is associated with the [\ion{N}{II}] emission and also visible in \ion{H}{I}, thus presumably associated with diffuse and ionized gas. It is also not traced by the low-J CO lines \citep{2015A&A...583A.107P,2015A&A...575A...9P}.\par

\begin{figure*}
   \centering
   \includegraphics[width=1\hsize]{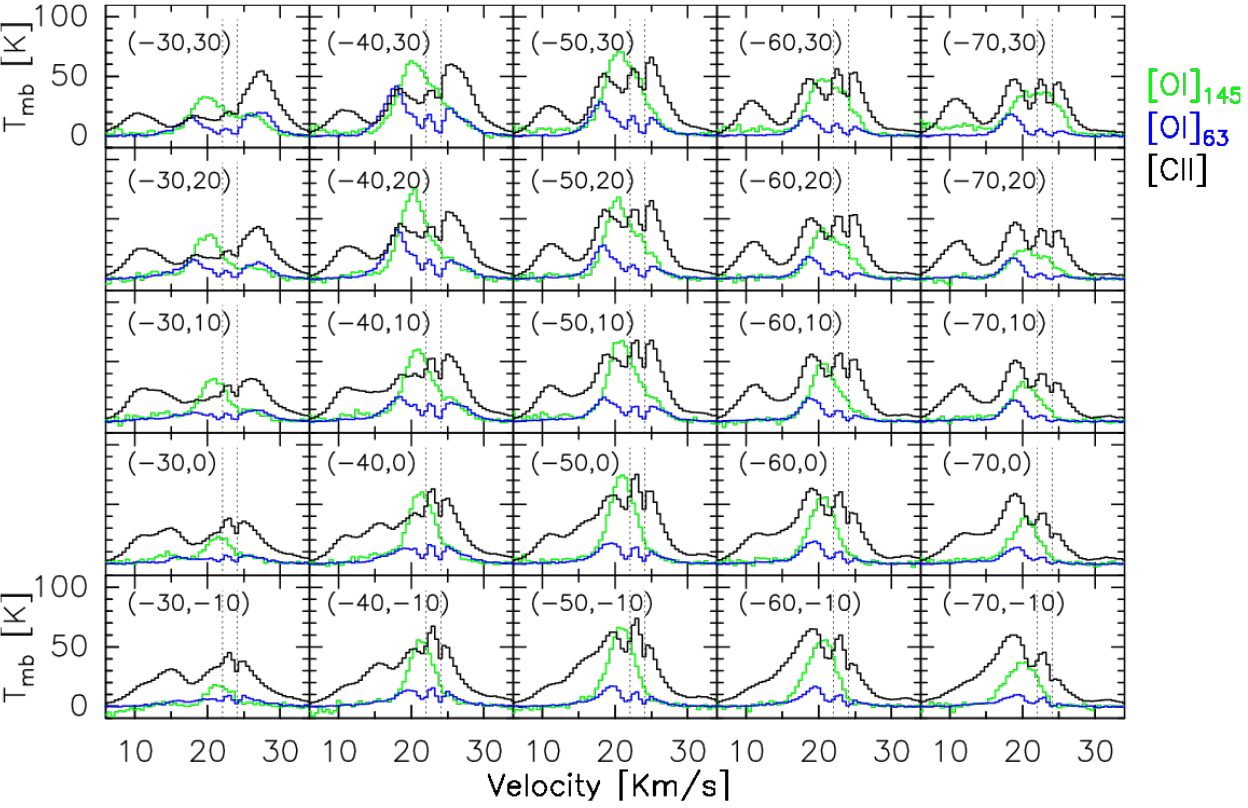}
      \caption{M17~SW \OI{63}, \OI{145}, and \Cp{} mosaic. The 
two dotted vertical lines per spectrum are located at 22 and 24~km s$^{-1}$, respectively, added to guide the eye to the absorption dips. The upper left  corner in each box gives the offsets of each spectrum.
              }
         \label{fig:M17OI63OI145}
   \end{figure*}

To check for positional variations of the \Oone~self-absorption profiles at small angular scales, we plotted a mosaic of the observed line profiles on a grid with 10\arcsec~separation, smoothed to the common spatial resolution of 15\arcsec. Figure~\ref{fig:M17OI63OI145} shows the spectra in \OI{63}, \OI{145} and \Cp{}. The \Cp{} spectra show spatial variations in their profiles from position to position, namely~across 10\arcsec, respectively, 0.10~pc, both in the absorption depth and the velocity of the absorption dips. The variations are similarly present in the \OI{63} spectra, although the deep absorption down to close to the zero intensity level makes the variations less pronounced in intensity. The velocities of the absorption dips are well correlated between the \OI{63} and the \Cp{} emission (same as above, see Sect.~\ref{foreground_physical}). In contrast, the \OI{145} shows a relatively simple profile of smoothly superimposed emission components. 
\par   
   
Table~\ref{table:OIintegrated} compares the peak and integrated intensities between both \Oone~line transitions for both sources. In general, the \OI{145} peak intensity overshoots the \OI{63} peak intensity by a factor of a few due to the self-absorption effect, except for M17~SW position 4. The latter shows a comparable peak intensity but at different velocities, again a result of self-absorption. The difference is more evident in the integrated intensity, with relatively greater \OI{145} integrated intensities, giving a 145/63 integrated intensity ratio between 1.2 and 7. The self-absorption effects thus significantly enhance the integrated intensity ratio. This effect must be considered when analyzing data, particularly non-resolved data without velocity information, such as high-redshift extragalactic observations. Although this self-absorption effect has already been observed under different environments of Galactic molecular clouds \citep[such as][ and references there in]{leurini15,gusdorf17,kristensen17,2018A&A...617A..45S}, it has also been observed in the large scale \OI{63}~spectra of extragalactic sources such as Arp~220, NGC~4945, NGC~4418, and even in the [\ion{O}{III}] 88 \um~of IRAS17208- 0014 \citep[e.g. ][]{gonzalez-alfonso12,fernandez-ontiveros16}. \par

\begin{table*}
  \centering
  \caption{Mon~R2 and M17~SW \OI{145} and \OI{63} integrated and peak intensity.}
  \label{table:OIintegrated}
  \begin{threeparttable}
  \begin{tabular}{c | c c c | c c c }
\hline
\hline
Positions  & \multicolumn{3}{c|}{Peak intensity} & \multicolumn{3}{c}{Integrated intensity} \\
\hline
            & \Oone$_{145}$ & \Oone$_{63}$ & $\frac{\Oone_{145}}{\Oone_{63}}$ & \Oone$_{145}$ & \Oone$_{63}$ & $\frac{\Oone_{145}}{\Oone_{63}}$ \\
            & (K) & (K) &  & (K km s$^{-1}$) & (K km s$^{-1}$) & \\
\hline
MonR2 1 & 65 & 12 & 5.4 & 394  & 111 & 3.5 \\      
MonR2 2 & 85 & 53 & 1.6 & 495  & 228 & 2.2 \\
\hline
M17SW 0 & 60 & 23 & 2.6  & 260 & 119 & 2.2 \\
M17SW 1 & 41 & 6 & 6.8   & 197 & 30  & 6.6 \\
M17SW 2 & 42 & 4 & 10.5   & 233 & 33  & 7.1 \\
M17SW 3 & 26 & 18 & 1.4  & 199 & 77  & 2.6 \\
M17SW 4 & 71 & 71 & 1.0 & 421 & 363 & 1.2 \\
M17SW 5 & 11 & 7 & 1.6    & 74 & 46   & 1.7 \\
M17SW 6 & 66 & 27 & 2.4  & 339 & 129  & 2.6 \\
\hline
\end{tabular}
\end{threeparttable}
 \end{table*}

\section{Gaussian multi-component analysis} \label{multicomponent}

For a quantitative comparison, we utilized a two-layer radiative transfer model (Gaussian multi-component fit to the \Oone~profiles) that is similar to what we introduced in our previous \Cp{12} and \Cp{13} study \citep{2020A&A...636A..16G}. The formal solution of the radiative transfer equation for a two-layer model with $b=1\ldots B$ number of background components and $f=1\ldots F$ number of foreground components gives for the brightness temperature of the line transition $t$ ($t=145,63$ denoting the \OI{145} and the \OI{63} transition respectively):

 \begin{equation}  \label{eq:tmb} 
\begin{split}
  T_{\mathrm{mb},t}(v) = & \left[ 
\mathcal{J}_{\nu}(T_{\mathrm{ex}_{t}}) \, \left( 1-\exp\left(-\sum_{b}\tau_{b,t}(v)\right)\right) 
\right] 
\exp\left( -\sum_{f} \tau_{f,t}(v)\right) + \\
 & 
\mathcal{J}_{\nu}(T_{\mathrm{ex}_{t}}) \, \left( 1-\exp\left( -\sum_{f} \tau_{f,t}(v)\right)\right).  \\
\end{split}
\end{equation}

Here, $\mathcal{J}_{\nu}(T) = \frac{h\nu}{k}\left(e^{h\nu/k_{\mathrm{b}}T}-1\right)^{-1}$ is the expression for the equivalent brightness temperature of a blackbody emission at temperature, $T$. The optical depth of each transition, $t,$ and component $i=b,f$ is given by: 

\begin{equation}
\tau_{i,t}(v)= \frac{c^3}{8\pi\nu_t^3} \phi_{i}(v) N_{u_t,i}\, A_t\,  
\left( \exp\left(h\nu_t/(k_b\,T_{\mathrm{ex}_{i,t}})\right)-1 \right),
\label{eq:tauNu4}
\end{equation}
with the upper and lower state of transition $t$ given by $u_t$ and $l_t$ respectively,  its upper state column density by $N_{\mathrm{u}_t,i}$, and its Einstein-$A$-coefficients, $A_t=A_{u_t,l_t}$. The numerical values of the latter are $A_{145}$ = 1.75$\times$10$^{-5}$ and $A_{63}$ = 8.91$\times$10$^{-5}$~s$^{-1}$ \citep{2007JPCRD..36.1287W}. The line profile of each component is: 

\begin{equation}
\phi_i(v) = \left( \frac{4 \ln(2)}{\pi} \right)^{1/2} \frac{1}{\Delta v_{\mathrm{FWHM},i}} \exp \left( {\frac{-(v-v_{\mathrm{LSR},i})^2 4 ln(2)}{\Delta v_{\mathrm{FWHM},i}^2}}  \right)
\label{eq:phi2}
,\end{equation} 
where $v_{\mathrm{LSR},i}$ the local standard of rest (LSR) velocity and $\Delta v_{\mathrm{FWHM},i}$ the full-width-half-maximum velocity line width of the line from component $i$. The upper and lower state column densities of a particular transition are related through the Boltzmann equation by the excitation temperature of the transition:

\begin{equation}
\label{eq:Boltz}
\frac{N_{u_{t,i}}}{N_{l_{t,i}}} = \frac{g_{\mathrm{u_t},i}}{g_{\mathrm{l_t},i}}\exp \left(\frac{\Delta E_{t,i}}{k_{\mathrm{b}}T_{\mathrm{ex}_t}} \right)
,\end{equation}
where $\Delta E_t = h\,\nu_t$. The optical depth of transition $t$ and component $i$ can thus be alternatively expressed by its lower state column density as:

\begin{equation}
\tau_{i,t}(v)= \frac{c^3}{8\pi\nu_{t}^3} \phi_{i}(v) \frac{g_{u_t}}{g_{l_t}} N_{l_t,i}\, A_t  \left(1 - \exp\left( -h\nu_{t}/k_b\,T_{\mathrm{ex}_{i,t}}\right) \right)
\label{tauNl3}
.\end{equation}

The parameters describing the line ($t=145,63$) intensity of each background ($i=b$) and foreground ($i=f$) component are thus four parameters: its excitation temperature, $T_{\mathrm{ex}_{i,t}}$, either its upper-state, $N_{u_t,i}$, or lower state, $N_{l_t,i}$ column density, and its line width, $\Delta v_{\mathrm{FWHM},i}$ and its line center position, $v_{\mathrm{LSR},i}$. \par

We assume that the excitation temperatures of all background components is the same (sufficient to obtain good fits, see below), so that we have one common value for the excitation temperature of the \OI{145} transition: $T_{\mathrm{ex}_{b,145}} \equiv T_{\mathrm{ex}_{145,bg}}; \, b=1\ldots,B$ and another common one for the \OI{63} transition: $T_{\mathrm{ex}_{b,63}} \equiv T_{\mathrm{ex}_{63,bg}};\, b=1\ldots B$, and correspondingly for the \OI{63} foreground lines. We note that for the three-level \Oone~system, the upper-state column density of \OI{63} is the same as the lower-state column density of the \OI{145} line. Thus, three parameters, namely, the column density in any of the three states (or the total column density) and the two excitation temperatures of both transitions, are sufficient to describe each component completely.\par 

\subsection{Fitting the background emission}\label{backfit}

The observed line profiles are then fitted by a two-layer model with a background in emission ($b$) composed of several components ($N_b$) and an absorbing foreground ($f$) with a different number of components ($N_f$), each with the fit parameters above. As we do not know the details of the excitation of the three levels (two excitation temperatures) of the atomic oxygen fine structure level system, we  used the peak main beam brightness temperature of \OI{63}, shining through in between the absorption notches, for the estimation of the background excitation temperature of \OI{63} by applying the inverse Rayleigh-Jeans correction. Then, we derived the background \OI{145} excitation temperature via the analytic solution to the balance equation between collisional excitation and de-excitation with spontaneous emission. We refer to Appendix~\ref{Appendix:tex} for a detailed  description of the procedure. The outcome is almost independent of the assumed density. The resulting kinetic temperatures in the background layer are given in Table~\ref{table:Tex_Tkin} for a density of $10^{6}$~cm$^{-3}$. In Sect.~\ref{Discussion} we further investigate how sensitive the derived column densities of the background and foreground layers are against variations of the assumed density and kinetic temperature of the background layer.

\begin{table}
 \centering
  \caption{Mon~R2 and M17~SW background excitation temperatures for both transitions and their respective kinetic temperatures.}
  \label{table:Tex_Tkin}
  \begin{threeparttable} 
  \begin{tabular}{c|c c c}
      \hline
\hline
Positions & $T_{\mathrm{ex}_{b,63}}$\tnote{a} & $T_{\mathrm{ex}_{b,145}}$\tnote{b} & T$_{\mathrm{kin}}$\tnote{b} \\ 
 & (K) & (K) & (K)\\
         \hline
    MonR2 1 & 78 & 144 & 90 \\      
    MonR2 2 & 140 & 227 & 152 \\
       \hline
    M17SW 0 & 95 & 200 & 114 \\
    M17SW 1 & 65 & 113 & 74 \\
    M17SW 2 & 70 & 125 & 80 \\
    M17SW 3 & 86 & 170 & 102 \\
    M17SW 4 & 145 & 240 & 158\\
    M17SW 5 & 67 & 117 & 77\\
    M17SW 6 & 100 & 222 & 122\\
        \hline
  \end{tabular}
  \begin{tablenotes}
    \footnotesize
  \item[a] RJ-corrected peak $T_{\mathrm{mb}_{b,63}}$ shining through.
  \item[b] Derived from 3-level rate equation solution, see Appendix~\ref{Appendix:tex}, at $n_{H_2}=10^6\rm{cm}^{-3}$.
  \end{tablenotes}
  \end{threeparttable}
\end{table}

The  kinetic temperatures for the atomic oxygen  derived in this way are in the temperature regime expected for PDR-gas, heated by the photo-electric effect well beyond the substantially lower dust temperature in the PDR material. For the dust, \citet{2017A&A...607A..22R} and Schneider N. (priv. comm.) have derived for Mon~R2 and M17~SW, respectively, dust temperatures below 30~K (see a comparison between gas and dust results in Sect.~\ref{gasanddust}). The \element{C}\element{O} rotational lines, particularly in the mid-J transitions, trace warm molecular gas from the PDR in a similar temperature range as we have derived here for atomic oxygen. Higher kinetic temperatures than the ones given in Table~\ref{table:Tex_Tkin} are possible; their impact on the derived column density of the background is explored in Sect.~\ref{backforetex}. \par

We started with the fit of the observed \OI{145} line profiles. First, we fit the background emission in the \OI{145} line profile by several emission line background components composed of a Gaussian optical depth profile scaled by the R-J correction. The excitation temperature was fixed to the common value, $T_{\mathrm{ex}_{b,145}} \equiv T_{\mathrm{ex}_{145,bg}}$, as discussed above. The fit parameters are the upper state column density of each background component $b$, $N_{u_{145},b}$, their LSR velocity $v_{\mathrm{LSR},b}$, and their line width $\Delta v_{\mathrm{FWHM},b}$, using the lowest possible number of Gaussian components. We note that a higher number of Gaussian components will not increase the total column density. Thus, there is no reason to use more components than needed. The process is iterative and we increased the number of components one by one until there is no substantial decrease in the chi-square of the fit with additional components. Then, we stopped the iteration. In this way, the simple, nearly Gaussian profiles of \OI{145} require one or two components, $B=B_{145}$, for the fitting.\par

Next, we identified the contribution of these $B_{145}$ components to the line wing emission of the lower transition, \OI{63}. The \OI{63} upper-level column density is fixed at the value derived from the upper state column density of the \OI{145} line, converted using the excitation temperature of the background via the Boltzmann-relation (Eq.~\ref{eq:Boltz}). The line center position, $v_{\mathrm{LSR},b}$, is also fixed within a narrow range (10\% of the value) for each component, namely to the value fitted for the \OI{145} profile, and the \OI{63} excitation temperature is fixed to the value of $T_{\mathrm{ex}_{63,bg}}$ derived above. The fit results show that the fitted width is typically less than 10\% smaller for the \OI{63} line, and thus consistent with the width fitted for the upper line. The fitting is restricted to the \OI{63} line wings not affected by foreground absorption; the relevant velocity ranges are from 2 to 5 and 14 to 18~km s$^{-1}$ for Mon~R2 and from 12 to 15 and 25 to 28~km s$^{-1}$ for M17S~SW respectively.\par

These fitted \OI{63} line profiles, resulting from the $b=1 \ldots B_{145}$ background components, show (outside of the core emission, which is heavily blended with the self-absorption) weak fit residuals requiring $B_{63}$ additional background velocity components, which we number by $b=B_{145}+1,\ldots, B_{145}+B_{63}$. We fit these by fixing the excitation temperature to the value used for the background emission. However, we can allow the column densities of the additional \OI{63} emission components, $N_{u_{63},b}$, to vary freely, as well as their line widths and central velocities, $\Delta v_{\mathrm{FWHM},b}, \,v_{\mathrm{LSR},b}$. These components are not visible in \OI{145} due to their low column density derived from the fit. This effect can be verified by calculating the corresponding \OI{145} background emission, which turns out to be of the order of 0.1~K or lower. We followed a step-by-step procedure instead of simultaneously fitting both \Oone~lines because the weak background components are only visible in \OI{63}, while the main components come from \OI{145}. Therefore, we did not expect a one-to-one correlation between the two \Oone~lines. \par

The total column densities of oxygen are listed in Table~\ref{table:OIbacktable}, obtained from the upper and lower \Oone~transitions derived from fitting the background emission profiles for the \OI{145} transition (background only) and the \OI{63} transition. These column densities are obtained by adding up all velocity components; however, the total column density is dominated by the one to two components $b=1\ldots B_{145}$ dominating the \OI{145} emission. We note that the detailed fit parameters of all components and levels are listed in \href{https://zenodo.org/doi/10.5281/zenodo.13800535}{DOI:10.5281/zenodo.13800536
}. \par

\subsection{Fitting of the foreground component}

In the next step, we dealt with the foreground absorption visible in the \OI{63} line. The column densities of the foreground absorption components are derived by fitting the absorption profiles against the fitted line profile of the \OI{63} background emission. The intensity in the self-absorption notches drops to very low values. This intensity requires correspondingly low \OI{63} excitation temperatures of the absorbing material. Due to the high energy of the \OI{63} transition, the resulting optical depth hardly depends on the exact value of $T_{\mathrm{ex}}$. As we do not have a way to independently derive a value of the excitation temperature of the foreground material, we fix the foreground excitation temperature to a single and low value of $T_{\mathrm{ex}_{f}} =$ 20~K. The population of the \OI{63} upper level is very small at these low excitation temperatures, so the total column density is also very insensitive to the assumed, fixed $T_{\mathrm{ex}_{f}}$. In Sect.~\ref{backforetex} we verify this weak sensitivity of the fitted foreground column density on the exact value of the assumed value of $T_{\mathrm{ex}_{f}}$. Because of the low excitation temperature of the \OI{63} transition, the \OI{63} upper level, providing the lower level of the \OI{145} transition, is hardly populated so that the foreground does not contribute to the observed \OI{145} line profile at all; neither in emission nor in absorption, independent of its assumed excitation temperature. Therefore, the foreground fit is done to the \OI{63} profile alone. This approach is backed up by the lack of self-absorption nodges in the observed \OI{145} line profile.\par

\begin{figure}
\centering
\begin{subfigure}{1.0\hsize}
   \centering
   \includegraphics[width=1\hsize]{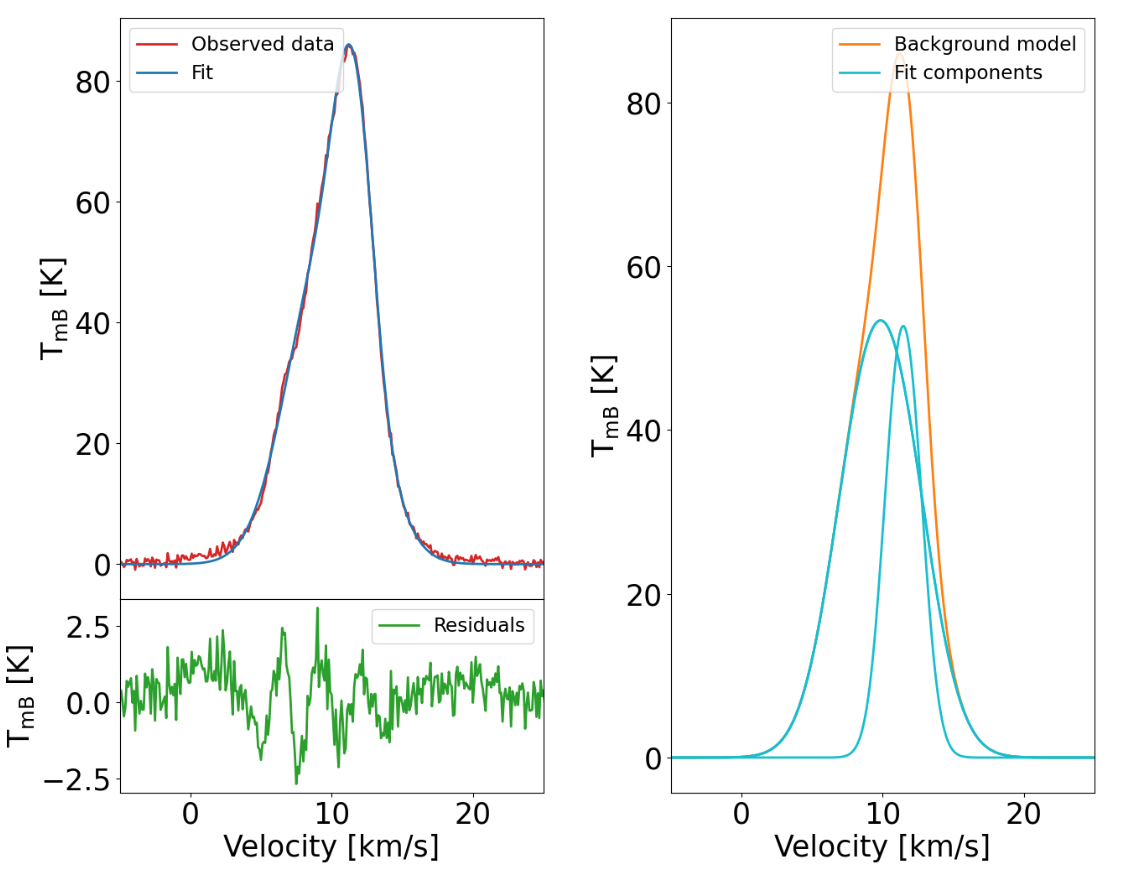}
         \label{fig:MonR2OI145}
\end{subfigure}%
\par
\begin{subfigure}{1.0\hsize}
   \centering
  \includegraphics[width=1\hsize]{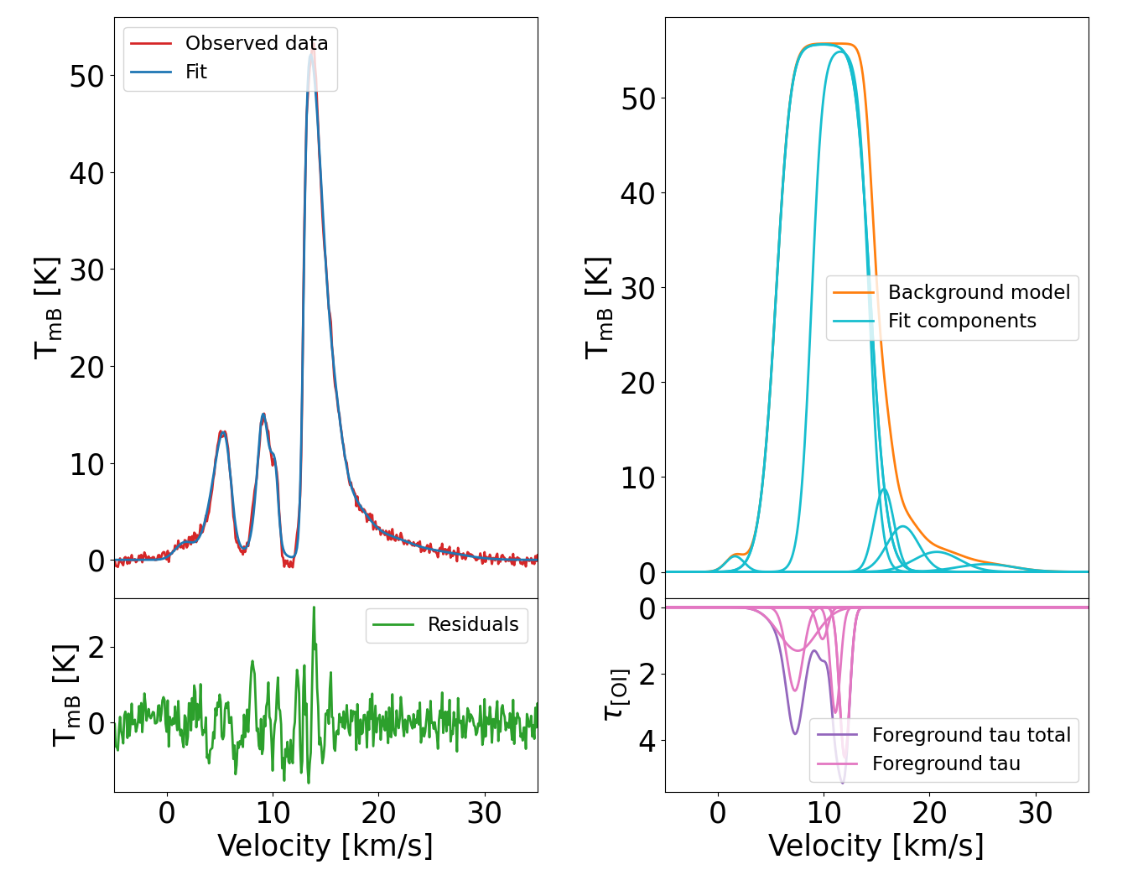}
      \label{fig:MonR2OI63}
\end{subfigure}
\caption{Mon~R2 \Oone~fits for position 2 for both transitions. The background fit needs 7 components, the 2 coming from the \OI{145} line carrying the bulk of the column density, plus 5 additional weak ones. The foreground fit needs 6 components. \textit{Top:} Mon~R2 \OI{145} fitting for position 2. \textit{Top left}: This panel represents the fitted model in blue and the observed spectra in red. \textit{Bottom left}: Residual curve between the observed spectra and the model. \textit{Right}:  Composition of each fitted component in cyan. \textit{Bottom:} Mon~R2 \OI{63} fitting for position 2. \textit{Top left}: Fitted model in blue and the observed spectra in red. \textit{Bottom left}: Residual curve between the observed spectra and the model.\textit{ Top right}: Composition of each fitted component in cyan and the fitted \OI{63} background profile in orange. \textit{Bottom right}: Foreground optical depth for the \OI{63} line of each Gaussian component in pink.}
\label{fig:MONR2fitting}
\end{figure}

With the excitation temperature fixed to this low value, this leaves the foreground column densities, $N_{l_{63},f}$, line width, $\Delta v_{\mathrm{FWHM},f}$, and line center, $v_{\mathrm{LSR},f}$, as the fitting parameters of the foreground components. As the \Cp{} and \Oone~absorption dips are very similar in line center velocity and width, we use the line center position and width of the \Cp{} foreground components fitted by \citet{2020A&A...636A..16G} as initial guesses for the foreground \Oone~fitting of the most substantial absorption components. We keep the velocity of the \Cp{} components fixed, allowing for a minor variation in width, and we freely vary the oxygen column density. An example of the fitting is shown in Fig.~\ref{fig:MONR2fitting}, which shows the results for position 2 in Mon~R2 for both \Oone~lines. The thus fitted column densities of the foreground are listed in Table~\ref{table:OIforetable}.\par

We note that a few Gaussian components have narrow line widths below 1~km~s$^{-1}$. These narrow components are located at the wings of the other narrow absorption dips. They are necessary for the fit to match the non-Gaussian absorption profiles. We do not regard them as independent physical components. They all have small amplitudes and, thus, we added their correspondingly small column densities to the column densities of the adjacent components. \par

\begin{table}
  \centering
  \caption{ Mon~R2 and M17~SW \OI{145} and \OI{63} background column density per position for the two layer model.}
  \label{table:OIbacktable}
  \begin{threeparttable}
  \begin{tabular}{c | c c | c c | c}
\hline
\hline
  & \multicolumn{2}{c}{\Oone$_{145}$} &  \multicolumn{2}{|c|}{\Oone$_{63}$}\\
  \hline
Positions  & \#   &    $\tau_{bg}^{*}$\tnote{a} & \#  &  $\tau_{bg}^{*}$\tnote{a} & $N_{bg}$(\element[0][]{O})\tnote{b}   \\
            & Back.        &          &  Back.  &       &     \\
            & Comp.  &    & Comp.   &       & (cm$^{-2}$)  \\
\hline
MonR2 1    & 3  & 0.8  & 7  & 42.1 & 5.1E+19 \\     
MonR2 2    & 2  & 0.3  & 7  & 6.3 & 1.3E+19 \\
\hline
M17SW 0 & 3  & 0.3 & 7 &  11.1 & 1.3E+19 \\
M17SW 1 & 1  & 0.8 & 2 &  72.1 & 5.0E+19 \\
M17SW 2 & 1  & 0.7 & 2 & 48.5 & 3.9E+19 \\
M17SW 3 & 2  & 0.2 & 3 &  9.7 & 1.3E+19 \\
M17SW 4 & 2  & 0.3 & 3 &  5.4 & 9.8E+18 \\
M17SW 5 & 3  & 0.1 & 3 &  6.6 & 1.3E+19 \\
M17SW 6 & 2  & 0.4 & 2 &  13.5 & 1.5E+19 \\
\hline
\end{tabular}
\begin{tablenotes}\footnotesize
\item[a] $\tau^*$ corresponds to the peak optical depth of the peak component, calculated from the fitted parameters according to eq.~\ref{eq:tauNu4}.
\item[b] Total column density, calculated from the sum of the three levels.
\end{tablenotes}
\end{threeparttable}
 \end{table}
 
 \begin{table}
  \centering
  \caption{Mon~R2 and M17~SW \OI{63} foreground column density per position for the two-layer model.}
  \label{table:OIforetable}
  \begin{threeparttable}
  \begin{tabular}{c | c  c  c }
      \hline
\hline
Positions   & \# &   $\tau_{fg}^{*}$\tnote{a} & $N_{fg}$(\element[0][]{O})  \\
    &         Fore. &  &                         \\
     &  Comp. &            & (cm$^{-2}$)     \\
          \hline
    MonR2 1  & 4  & 12.1 & 4.6E+18 \\
    MonR2 2  & 6  & 5.4 & 3.9E+18 \\
       \hline
    M17SW 0  & 7 & 3.6& 1.8E+18  \\  
    M17SW 1  & 4 & 5.7 & 3.3E+18  \\
    M17SW 2  & 6 &  3.1 & 3.4E+18  \\
    M17SW 3  & 5 &  3.9 & 3.0E+18  \\
    M17SW 4  & 8 &  2.8 & 2.5E+18  \\
    M17SW 5  & 5 &  3.0 & 3.9E+18  \\
    M17SW 6  & 6 & 1.8 & 2.5E+18  \\
        \hline
\end{tabular}
\begin{tablenotes}\footnotesize
\item[a] Same as in Table~\ref{table:OIbacktable}, resp.~ Eq.~ \ref{eq:tauNu4}.
\item[b] Same as in Table~\ref{table:OIbacktable}.
\end{tablenotes}
\end{threeparttable}
 \end{table}

\subsection{Foreground and background column densities}\label{forebackcoldens}

The derived total column density of the \Oone~background and foreground, given by the sum of the individual velocity components, are listed in Tables~\ref{table:OICIItable_back} and \ref{table:OICIItable_fore}, respectively. For easier comparison, the column densities are also converted to hydrogen column densities, $N_{H}$, and to equivalent visual extinctions, assuming that all oxygen is in the form of \element[0][]{O}. From \citet[][their table 1, for high-metal elemental abundances]{2008ApJ...680..371W}, we use an \element{O}/\element{H} abundance ratio of 2.56 $\times$ 10$^{-4}$. We consider the canonical conversion factor between the total hydrogen column density and visual extinction of 1.87$\times$10$^{21}$~cm$^{-2}$/A$_{\rm V}$ \citep{1978ApJ...224..132B}. Tables \ref{table:OICIItable_back} and \ref{table:OICIItable_fore} also compare the total oxygen column densities of the foreground and background components to the ones derived previously for \Cp{}. The \element[+][]{C} column density also has been converted to an equivalent visual extinction using a \element{C}/\element{H} abundance value from \citet[][same as above, table 1]{2008ApJ...680..371W} of 1.2$\times$10$^{-4}$ and the canonical conversion factor listed above to convert to magnitudes. The tables also list the ratio of the \Oone~and \Cp{} column densities and the ratio of the equivalent extinctions. The abundance ratio between oxygen and carbon is 2.1 when using the elemental abundances relative to hydrogen quoted above.\par 

To ease the visualization of the multiple foreground and background components, both for the new \Oone~and the former \Cp{} data and, in particular, also for the discussion of the position-to-position variations below (see Sect.~\ref{foreground_physical}), we compared the fitted parameters for different components in the foreground and background; namely, their velocities, velocity widths, and column densities, in Figs.~\ref{fig:background_all} and \ref{fig:foreground_all}

\begin{table*}
  \centering
  \caption{Mon~R2 and M17~SW comparison between oxygen and ionized carbon column densities and equivalent extinctions for the background columns.}
  \label{table:OICIItable_back}
 \begin{threeparttable}
  \begin{tabular}{c|c c c c c c}
      \hline
\hline
  & \multicolumn{6}{c}{Background}  \\
\hline
Positions & $N_{bg}$(\element[0][]{O}) & $N_{bg}$(\element[+][]{C})\tnote{a}  & $\frac{{N} (\element[0][]{O})}{{N} (\element[+][]{C})}$  & A$_{\mathrm{V},bg}$\element[0][]{O}\tnote{b} & A$_{\mathrm{V},bg}$\element[+][]{C}\tnote{a} & $\frac{A_{\mathrm{V},bg} \element[0][]{O}}{A_{\mathrm{V},bg} \element[+][]{C}}$ \\ 
 & (cm$^{-2}$) & (cm$^{-2}$) &   & (mag.) & (mag.) & \\
         \hline
    MonR2 1 & 5.1E+19 & 4.2E18 & 12.1 & 106 & 19 & 5.6 \\      
    MonR2 2 & 1.3E+19 & 4.7E18 & 2.8 & 28 & 21 & 1.3  \\
       \hline
    M17SW 0 & 1.3E+19 & 9.2E18 & 1.4 & 28 & 41 & 0.7  \\
    M17SW 1 & 5.0E+19 & 8.0E18 & 6.3 & 105 & 36 & 2.9  \\
    M17SW 2 & 3.9E+19 & 5.6E18 & 3.8 & 82 & 25 & 1.8 \\
    M17SW 3 & 1.3E+19 & 4.4E18 & 7 & 28 & 20 & 1.4  \\
    M17SW 4 & 9.8E+18 & 7.6E18 & 1.3 & 20 & 34 & 0.6  \\
    M17SW 5 & 1.3E+19 & 3.0E18 & 4.3 & 27 & 13 & 2.1  \\
    M17SW 6 & 1.5E+19 & 7.7E18 & 1.9 & 31 & 34 & 0.9  \\
        \hline
\end{tabular}
\begin{tablenotes}\footnotesize
\item[a] \Cp{12} column densities from \citet{2020A&A...636A..16G}.
\item[b] Equivalent visual extinction for the \Oone~and \Cp{} column densities (factors for abundance and conversion from $N(\element{H})$ as discussed in the text), with $N(\element{H})$ = 1.87$\times$10$^{21}$~cm$^{-2}$ A$_{\rm V}$.
\end{tablenotes}
\end{threeparttable}
 \end{table*} 
 
 \begin{table*}
  \centering
  \caption{Mon~R2 and M17~SW comparison between oxygen and ionized carbon column densities and equivalent extinctions for the foreground columns.}
  \label{table:OICIItable_fore}
 \begin{threeparttable}
  \begin{tabular}{c|c c c c c c}
      \hline
\hline
  & \multicolumn{6}{c}{Foreground}\\
\hline
Positions & $N_{fg}$(\element[0][]{O}) & $N_{fg}$(\element[+][]{C})\tnote{a}  & $\frac{N(\element[0][]{O})}{N(\element[+][]{C})}$  & $A_{\mathrm{V},fg}$\element[0][]{O}\tnote{b} & $A_{\mathrm{V},fg}$\element[+][]{C}\tnote{a} & $\frac{A_{\mathrm{V},fg}\element[0][]{O}}{A_{\mathrm{V},fg}\element[+][]{C}}$ \\ 
 & (cm$^{-2}$) & (cm$^{-2}$) &   & (mag.) & (mag.) & \\
         \hline
    MonR2 1 &  4.6E+18 & 8.3E17 & 5.5 & 9.7 & 3.7 & 2.6 \\      
    MonR2 2 &  3.9E+18 & 6.4E17 & 5.9 & 8.1 & 2.9 & 2.8 \\
       \hline
    M17SW 0 & 1.8E+18 & 2.0E18 & 1.0 & 3.8 & 9.2 & 0.4 \\
    M17SW 1 & 3.3E+18 & 1.7E18 & 1.9 & 6.8 & 7.6 & 0.9 \\
    M17SW 2 & 3.4E+18 & 3.0E18 & 1.1 & 7.1 & 13 & 0.5 \\
    M17SW 3 & 3.0E+18 & 7.7E17 & 4.0 & 6.2 & 3.5 & 1.8 \\
    M17SW 4 & 2.5E+18 & 1.3E18 & 1.8 & 5.1 & 5.8 & 0.9 \\
    M17SW 5 & 3.9E+18 & 3.9E17 & 10 & 8.2 & 1.7 & 4.8 \\
    M17SW 6 & 2.5E+18 & 1.8E18 & 1.4 & 5.3 & 8.0 & 0.7 \\
        \hline
\end{tabular}
\begin{tablenotes}\footnotesize
\item[a] \Cp{12} column densities from \citet{2020A&A...636A..16G}.
\item[b] Equivalent visual extinction of the total \Oone~ and \Cp{} column densities (factors for abundance and conversion from $N(\element{H})$ as discussed in the text), with $N$(\element{H}) = 1.87$\times$10$^{21}$~cm$^{-2}$ A$_{\rm V}$.
\end{tablenotes}
\end{threeparttable}
 \end{table*} 

We notice that the background column density ratio of \Oone~to \Cp{} is close to the elemental abundance in most positions; correspondingly, the equivalent A$_{\rm V}$ ratio is close to unity. This ratio is expected if all material is fully excited. However, due to the higher critical density of the \Oone~transitions, it is likely that more \element[+][]{C} is traced in emission. Therefore, we would expect to measure a background column density ratio between oxygen and ionized carbon, where the elemental abundance is the upper limit. However, the value for Mon~R2, position 1, in the background layer, clearly sticks out; the other two cases for an \Oone~excess in the background layer well above the elemental ratio are positions 1 and 5 in M17~SW. We note that for these positions, we use a relatively low $T_{\mathrm{ex}_{63,bg}}$, derived from the relatively low peak brightness of the \OI{63} line. The latter is a lower limit, which holds in the optically thick case; higher values for $T_{\mathrm{ex}}$ are perfectly feasible and would result in a correspondingly lower \Oone~column density, as discussed in Sect.~\ref{backforetex}. On the other hand, positions 0, 4, and 6 for M17~SW have a lower oxygen column density than the other positions. As discussed above, these values could be an effect of a too-low density of the collision partners to excite the oxygen; see Sects.~\ref{backforetex} and \ref{app:densityeff} for an explanation of how density affects the results. \par 

For the foreground layer, we expect an opposite situation. In cold material, some atomic oxygen may be left in the overall molecular material where most carbon is bound in CO. Therefore, we should measure a \Oone~to \Cp{} abundance ratio higher than the elemental abundance in absorption. Mon~R2 presents the higher ratios for both sources. Given the likely scenarios for the structure of Mon~R2 (see Sect.~\ref{foreground_physical}), it is expected that this excess is due to the inherent structure of the source. M17~SW shows a different behavior. The values that deviate from what is expected are found in positions 0, 2, 4, and 6.\par

Interestingly, the affected positions are closer to the main ridge (see Fig.~\ref{fig:M17array}). It is clear that a particular spatial configuration plays a role here. However, with the foreground layer being invisible and only being detected through the absorption profile of bright atomic lines, it is hard to speculate without much information. Future studies over the whole map for the foreground layer in both lines could help to resolve this issue.

\section{Discussion} \label{Discussion}

We have estimated the background and foreground column densities for neutral atomic oxygen under the assumption of a lower limit for the excitation temperature in the background layer along the line of reasoning discussed in Appendix~\ref{Appendix:tex}. With the estimated range of excitation temperatures for \OI{63} between 65 and 145~K, we derive column densities from the fit to the observed line profiles between 8.5 and 49 $\times$ 10$^{18}$~cm$^{-2}$ for the background layer and between 1.8 and 4.6 $\times$ 10$^{18}$~cm$^{-2}$ for the foreground layer. In the following, we first evaluate how sensitively the results are affected by the necessary simplifying assumptions we made. Then, we turn to a discussion of the characteristics of the foreground layer.

\subsection{Effect of variations in the background and foreground physical parameters} \label{backforetex}

The choice of the value for the excitation temperature of the background layer in both \Oone~transitions is a critical element of the study, as the temperature in the foreground layer. As presented above, we have taken the excitation temperature of the \OI{63} transition in the background as the lower limit derived from the peak brightness temperature of this transition observed at each position; with the assumption of high density, necessary because of the high critical density of the \OI{145} transition, we then use the analytical solution for the population of the three-level \Oone~fine structure system (see Appendix~\ref{Appendix:tex}) to derive the excitation temperature for the upper transition (assuming $10^6$~cm$^3$ for the density). To estimate how sensitive the derived column densities for the background layer components are to these assumptions, we analyzed how the result depends on varying first the density (see Appendix~\ref{app:densityeff}), then  on the kinetic temperature of the background components (see Appendix~\ref{app:kinetic}).  \par 

This analysis shows that the resulting background column densities span a range in physical scenarios of warm PDR material with correlated temperatures; they can reach about ten times larger column densities for the background compared to the minimum derived for the nominal parameters. Thus, from this analysis alone, the amount of warm PDR material traced by the atomic oxygen background emission cannot be constrained further but is in a range entirely consistent with standard PDR scenarios and also with the range of column densities for warm PDR gas traced by mid- and high-J CO lines \citep[for example, ][]{1988ApJ...332..379S,2012A&A...544A.110P,2015A&A...583A.107P}.  \par

Finally, we vary the foreground excitation temperature and study its effects on the derived foreground column density (see Appendix~\ref{app:foresect}). The foreground excitation temperature analysis shows that the column density is not much affected by the assumed value for the foreground excitation temperature, changing monotonically by not more than 20\% over the possible temperature range. Thus, we conclude that the foreground column density of the absorption components, derived from the fit to the complex line profiles, is a robust result. \par

\subsection{Column densities of the different components of the background layer}

\begin{figure*}
   \centering
   \includegraphics[width=0.8\hsize]{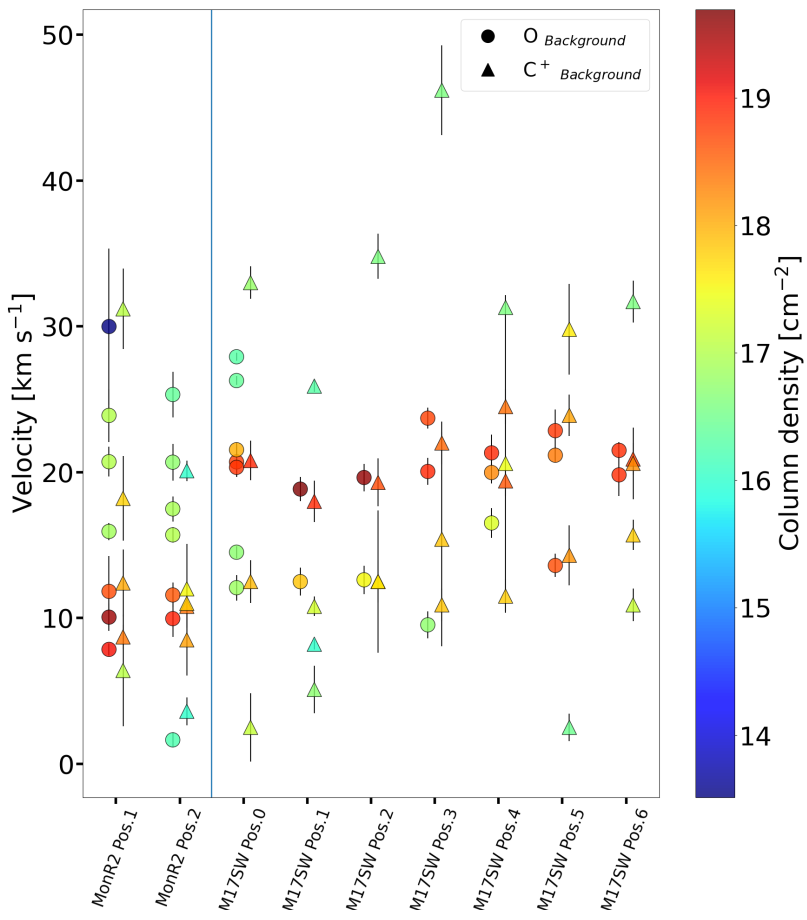}
      \caption{Compilation of background Gaussian parameters for all positions from the multi-component analysis from Sect.~\ref{Results}. Each position is plotted in columns along the velocity axis, showing both atomic lines. Circles symbolize the \Oone~Gaussian components for each position and triangles the \Cp{} ones \citep[from ][]{{2020A&A...636A..16G}}. The color of each symbol represents its column density. The length of the vertical bars denotes half of the line width value for a better visualization.
      }
         \label{fig:background_all}
   \end{figure*}

One of the central motivations behind this work is to study the nature of the foreground layer identified through the self-absorption traced by both cooling lines, \Cp{} and \Oone. Nevertheless, we also briefly discuss the nature of the background layer. In Fig.~\ref{fig:background_all}, we plot the column density for each fitted Gaussian velocity component for both \Cp{} and \Oone~for each position analyzed. \Cp{} is characterized by diamonds and \Oone~by circles. 

One or two components dominate the background column density contribution for both atoms around the systemic velocity, 10~km~s$^{-1}$ for Mon~R2 and 21~km~s$^{-1}$ for M17~SW with more than 80\% of contribution from these main components. We also see that the additional components tracing the line wings contribute only marginally to the column density, with Mon~R2 showing a red-shifted line wing and M17~SW showing emission at both sides skewed to the blue side of the spectra. \par

We notice that the main background emission, both in \Cp{} and in \Oone, is fitted by a few well-overlapping Gaussian components, resulting in a smooth overall profile. Sometimes even only one single component dominates, in particular in \Cp{}, where the bulk background emission is defined by the \Cp{13} line profile. The \Oone~background emission, having to match both the \OI{145} profile and the line wings of heavily self-absorbed \OI{63} profile, sometimes requires several more Gaussian components; these are, however heavily overlapping and sum-up to a similar total column density than the single \Cp{} component (one example is the three components at M17~SW pos.~0, or the two components in \Oone~matching the single component in \Cp{} at M17~SW pos.~3). \par
 
Nonetheless, the dominant contribution in column density comes from components where both are present and which have similar velocities and line-width in \Cp{} and \Oone. We can see that the background is not entirely uniform and shows some variations from position to position. To conclude, the background, at the different positions, shows a smooth line structure dominated by one or two Gaussian components and consistent velocities around the systemic velocity.

We also note that the center velocities and line widths of ionized carbon and atomic oxygen are similar to the emission observed in other tracers, such as atomic carbon and carbon monoxide. For Mon~R2, the emission spans a range between 0 and 20~km~s$^{-1}$, with a peak intensity around 10~km~s$^{-1}$, not different from what we see in other tracers such as CO \citep[e.g.][]{2012A&A...544A.110P}; however, the CO emission shows much more complex line profiles with additional foreground absorption. The same is the case for M17~SW if we compare the emission to observations in [\ion{C}{I}] or \element[][]{C}\element[][]{O}, in particular, for the optically thin lines \citet{2015A&A...583A.107P,2015A&A...575A...9P}, with emission between 10 and 30~km~s$^{-1}$ and their intensities peaking around 20 or 22~km~s$^{-1}$, plus additional components outside this velocity range mentioned above, only visible in \Cp{}.

\subsection{Physical properties of the foreground layer}
\label{foreground_physical}

The significant result from the previous observations and analysis \citep{2020A&A...636A..16G} of the \Cp{} self-absorption notches was that they require a large foreground column of \Cp{} with very low $T_\mathrm{ex}$ material ($\leq$ 20~K). The present study shows that the \OI{63} line also shows deep self-absorption notches, which generally agree well in terms of velocity and line width with the \Cp{} self-absorption when both are present and correspond to similar total column densities. \par

\begin{figure*}
   \centering
   \includegraphics[width=0.8\hsize]{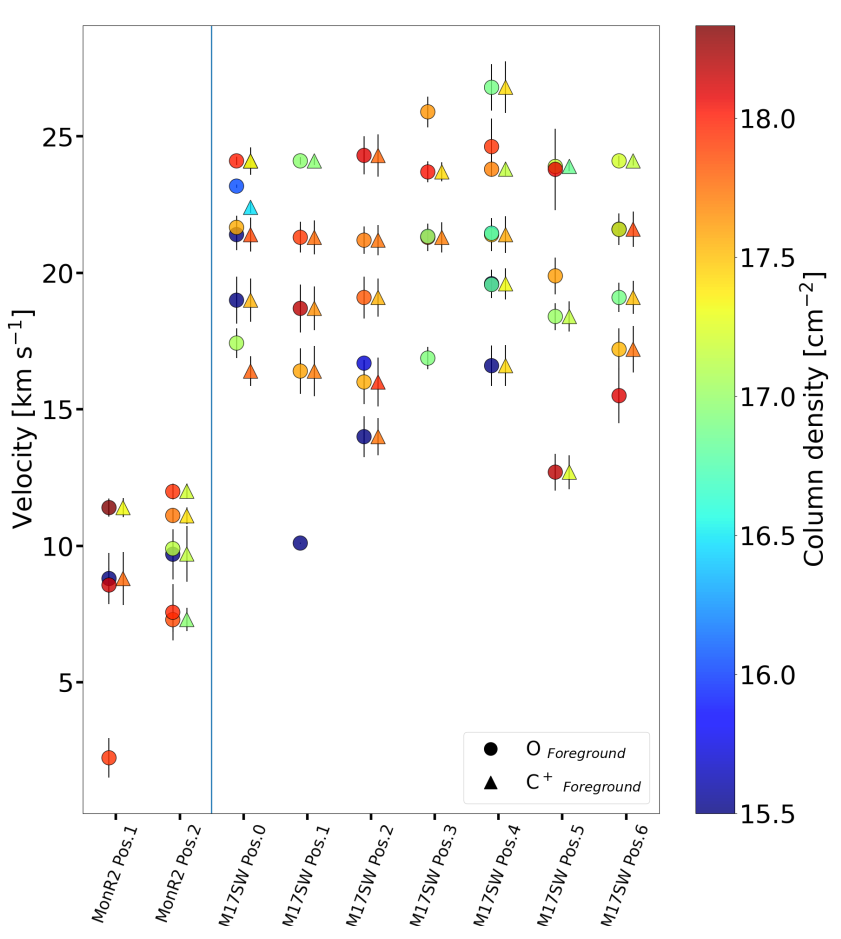}
        \caption{Same as Fig.~\ref{fig:background_all} for the foreground parameters.}
         \label{fig:foreground_all}
   \end{figure*}

Similarly to the background emission components analysis shown in Fig.~\ref{fig:background_all}, we display the column densities, velocities, and line widths of the individual foreground components in Fig.~\ref{fig:foreground_all}. The figure demonstrates that the self-absorption notches in \Cp{} and \OI{63} have similar velocities and narrow line widths. Moreover, the different components at the other positions can be grouped around common velocities, with 8 and 12~km~s$^{-1}$ for Mon~R2, and 16, 18, 21, and 24~km~s$^{-1}$ for M17~SW. However, the exact velocity for each component fluctuates around these average values, the magnitude of these fluctuations being on the order of the line widths. \par

The total column density at each position is distributed between different velocities, and the column density of a given velocity component varies from position to position. As discussed above, most velocity components in the foreground are seen in both species, \Oone~and \Cp{}, with similar fit parameters. In several cases, the foreground absorption, being fitted by a single component in one species, requires two velocity components with different widths in the other. Some examples are M17~SW, pos.~0, where the \OI{63} absorption at 24.1~km~s$^{-1}$ is fitted by a single component with a width of 0.6~km~s$^{-1}$, whereas the corresponding \Cp{} absorption at 24.1~km~s$^{-1}$ needs two components: one with a wider line width of 2~km~s$^{-1}$ and a narrower one with a width of also 0.6~km~s$^{-1}$. A similar situation is shown for the 21.3~km~s$^{-1}$ center velocity component in M17~SW, pos.~3, where the \Cp{} absorption is fitted by a single component of width 2.2~km~s$^{-1}$ (Table~F.4 in \citet{2020A&A...636A..16G}) and the \Oone~absorption needs two fit components, one with width 2~km~s$^{-1}$, the other one with a with of 1.1~km~s$^{-1}$. These differences are presumably due to the oversimplifying assumption of purely Gaussian absorption line profiles; the different fit components, nevertheless, can safely be counted as contributions to the same absorption feature.\par

It is solely on very few occasions that an absorption component appeared only in \Oone and not in \Cp{}. This phenomenon only occurs in the line wings and for features with relatively low column density; examples are the 25.9~km~s$^{-1}$ and the 16.9~km~s$^{-1}$ component in M17~SW, pos.~3, or the 15.5~km~s$^{-1}$ component in M17~SW, pos.~6. In all these cases, the significance of the absorbing foreground fit components may be spurious, as the absorption is fitted against the weak and somewhat noisy line wings of the background \OI{145} emission line profiles. \par

In contrast, there are several cases where the \Cp{} line shows clear absorption components but where the corresponding \OI{63} absorption shows much lower column densities or is almost absent. This absorption is typically the case for the \Cp{} absorption on the low-velocity side of the line profiles in M17~SW. Examples are the 16.4 and 19.1~km~s$^{-1}$ components at pos.~0, see Table~F4 in \citet{2020A&A...636A..16G}, and the corresponding \OI{63} absorption component at 17.4~km~s$^{-1}$, or the 16.6 and 19.9~km~s$^{-1}$ \Cp{} components in pos.~4 (see Table~F4 as above) and the corresponding \OI{63} component at 19.6~km~s$^{-1}$. As the \Oone~emission does not show the weak wing emission below 17~km~s$^{-1}$, these features cannot show up in absorption in \OI{63}, even if they might be present in the foreground, as they lack background emission to absorb against.\par 

Both cases discussed above occur in the wings of the main line emission. The quantitative comparison of \Oone~and \Cp{} column densities in these components might be affected by the slightly different form of the emission line profiles from the background. Similarly, the quantitative comparison of the absorbing column densities from position to position, as discussed in the next paragraph, will be affected by the features in the line wings by corresponding variations in the line profile of the background emission.

\subsubsection{Spatial variations and density estimate}
\label{denstiy_estimate}

We can obtain a lower limit estimate for the density of the foreground material from the variations in the derived foreground column density between adjacent positions. To illustrate the range of densities and spatial scales involved, let us, for the moment, assume a diffuse foreground of density 100~cm$^{-3}$, already relatively high for a diffuse cloud. The atomic oxygen column densities derived for the foreground of around $2\times10^{18}{\rm cm}^{-3}$ (Table~\ref{table:OICIItable_fore}) together with the oxygen abundance quoted above would then imply a spatial extent along the line of sight of $7.8\times10^{19}{\rm cm}$. Taking this as the typical size of the assumed homogeneous foreground cloud would imply an angular extent of 45 arcmin ($\sim$30 pc) at the distance of M17~SW.\par

Based on the larger uncertainties for the derived column density values of the foreground absorption against the background line wings, as discussed above, we focus on the analysis of the variation in column density of the foreground components from position to position, on the absorbing components in the line cores, which are less affected by the uncertainties in the background brightness, namely the features at velocities between 18 to 25~km~s$^{-1}$. Comparing the absorption feature near 24.1~km~s$^{-1}$¸ at pos.~0 and pos.~1 of M17~SW, namely,~between a position on the bright interface ridge to one further into the molecular cloud, which has a total \ion{H}{I}-equivalent column density of $2.8\times 10^{21}~{\rm cm}^{-2}$ at pos.~0 and drops to $0.8\times 10^{21}~{\rm cm}^{-2}$ at pos.~1; hence, it shows a decrease by $2.0\times 10^{21}~{\rm cm}^{-2}$  in \Cp{}, and from $4.5\times 10^{21}~{\rm cm}^{-2}$ to $0.6\times 10^{21}~{\rm cm}^{-2}$, and, therefore, a decrease of $3.9\times 10^{21}~{\rm cm}^{-2}$  in \Oone. Comparing the column density variations between positions along the ridge (i.e~pos.~3 and pos.~0), the velocity components at about 21.3 to 21.7~km~s$^{-1}$¸ drop in \Oone~from ($3.5+0.6 \times 10^{21}~{\rm cm}^{-2}$) to $2.0\times 10^{21}~{\rm cm}^{-2}$ (i.e.,~ a  2.1 drop), whereas it increases from 4.2 to 6.2, namely,~an increase of 2.0 in \Cp{}. The rise in \Cp{} versus the decrease in \Oone~in this case indicates a change in the chemical composition along the line of sight from position to position. Repeating this comparison for all pairs of adjacent positions, we see that the variations from position, both in \Oone~and \Cp{}, are typically of the order of 1 to $2\times 10^{21}~{\rm cm}^{-2}$ and peak variations are up to about $4\times 10^{21}~{\rm cm}^{-2}$. This comparison includes the variations in column density between the two positions measured in Mon~R2.
 
\begin{figure}
\centering
   \includegraphics[width=1.0\hsize]{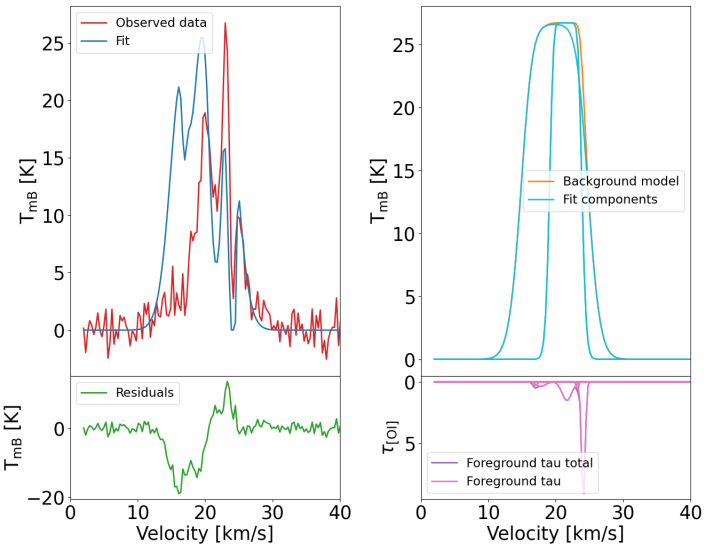}
      \caption{Unsuccessful (note the significantly increased level of the residuals) fit for M17~SW pos. 6, using the foreground parameters fitted to position 0.}
         \label{fig:M17fixed_1}
\end{figure}

\begin{figure}
   \centering
   \includegraphics[width=1.0\hsize]{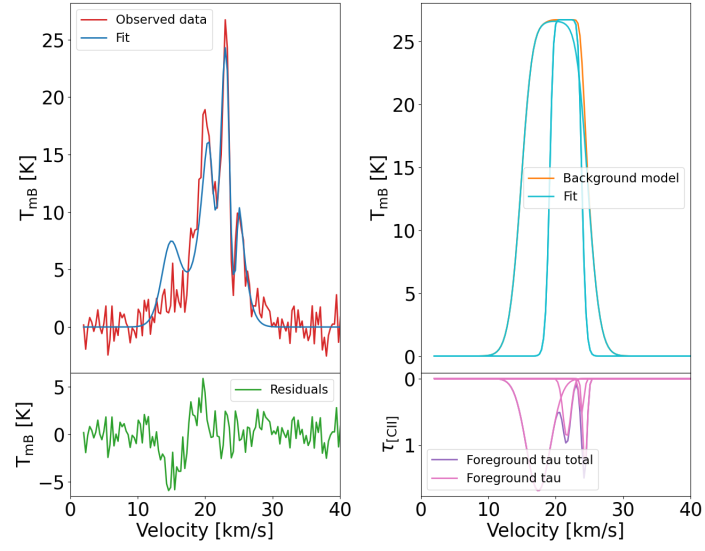}
      \caption{Same as Fig.~\ref{fig:M17fixed_1}, using the foreground parameters fitted to position 0, but allowing for variations in width and column density.}
         \label{fig:M17fixed_2}
\end{figure}

To understand to what degree the variations in the absorbing column densities from position to position might be an artifact of the ambiguity of the foreground absorption fit with several Gaussian components, we checked how the resulting fitted line profile looks like if we fix the foreground components to the parameters from the adjacent position, first in all three foreground line parameters: line center, width, and column density. We have done this test for position 1 in Mon~R2 with the parameters of position 2 and position 6 in M17~SW with the parameters of position 0. As an example, Fig.~\ref{fig:M17fixed_1} shows the resulting fit for M17~SW position 6. In a second test, we allowed each foreground component's column density and line width to vary, keeping the velocity (and excitation temperature) fixed (Fig.~\ref{fig:M17fixed_2} shows an example of the resulting fit). Both attempts are unsuccessful, as demonstrated by the significant residuals well above the noise level, with (obviously) slightly better results, allowing for more variations. Considering that the foreground absorption is multiplicative, $e^{-\tau_{fg}}$, implying that the optical depth is determined as a proportion of the background intensity and hence independent of its actual intensity and background intensity variation, this demonstrates that the variation in column density, and therefore optical depth, between adjacent positions derived from the complete fits of the two-layer model as discussed above, is a significant result. Keeping the column density constant across adjacent positions, as would be the case for a smooth and homogeneous foreground component, contradicts the observed line profiles.

Thus, we derived the variations in \ion{\element[][]{H}}{I} equivalent column density for each species to be between 1-2 and up to 4$\times~10^{21}~{\rm cm}^{-2}$. These are variations of the total column density, such as a decrease or increase in both species, or they indicate the variation of the chemical composition along each line of sight, in case the variation is in the opposite sense for both species. The linear scales between adjacent positions correspond to 0.27~pc at the distance of M17~SW and 0.12~pc for Mon~R2. \par 

As discussed in \citet{2020A&A...636A..16G}, assuming that the components observed are not all pencil-like directed along the line of sight, the magnitude of the column density variation, combined with the linear scale corresponding to the angular separation at the distance of the source, gives a constraint on the variation of the density of the foreground material and, hence, a lower limit estimate of the density of the foreground material. With the numbers quoted above, this results in minimum hydrogen volume densities of 1.2 to $2.4\times~10^{3}~{\rm cm}^{-3}$, maximum about $5\times~10^{3}~{\rm cm}^{-3}$ for M17~SW, and correspondingly higher, according to the closer distance and hence smaller linear scale between the observed positions, namely,~ 2.8 to $5.6\times~10^{3}~{\rm cm}^{-3}$, for Mon~R2.\par

We emphasize that these densities are a lower limit due to several factors: first, the densities are derived from beam-averaged column densities. Thus, if the source has structure on scales smaller than the beam, the column densities, and hence the derived densities, will be accordingly higher. Secondly, the densities are derived from the smallest observed distance, the pixel spacing of the upGREAT array; similar changes on smaller scales, as suggested by the profile changes shown for the fully sampled \Cp{} map and the regridded \Oone~map in Fig.~\ref{fig:M17OI63OI145} would give correspondingly larger densities. In addition, the densities can be higher if the absorbing material is a surface layer significantly smaller than the spatial size of the variations. And lastly, the derivation of the \ion{\element[][]{H}}{I} column densities assumes that all the material is in the form of \element{O}$^0$ and \element{C}$^+$; they would be correspondingly higher, and similarly for the volume densities, if only a fraction of the material is in the form of \element{O}$^0$ or \element{C}$^+$, a possibility that is already indicated by the variation of the \element{O}$^0$ and \element{C}$^+$ column density ratios for the different components and positions. The   lower limit for the density of the absorbing foreground derived in this way, estimated from the variation in column density from position to position, is two to three orders of magnitude greater than typical for homogeneous diffuse absorbing foreground layers. \par 

As a side note, we can estimate the kinetic temperatures for these densities that would result in the excitation temperature we have used for our analysis, 20~K, from the excitation temperature plots in Appendix~\ref{Appendix:tex}. This estimation gives a foreground kinetic temperature of 30~K for Mon~R2 and 38~K for M17~SW. The total column density of the low $T_\mathrm{ex}$, absorbing foreground, is very large as shown in Table~\ref{table:OICIItable_fore}, corresponding to an equivalent visual extinction of several $A_V$ or more. The agreement in velocity implies that it is associated with the source. Based on the arguments above about the spatial (in projection) variation of the derived column densities, we estimate it to have a relatively high density. Both the column and the volume densities are lower limits, assuming that all its carbon is in the form of \element[+][]{C} and all its oxygen is in the form of \element[0][]{O}. These arguments strongly suggest that the material is not a diffuse, smooth foreground layer but is directly associated with the background material, namely, the dense, strongly UV-illuminated PDR material.

The nature and origin of this material are somewhat puzzling: we would need a mechanism that produces such a large column of low $T_\mathrm{ex}$, dense material, but where the carbon stays ionized although the high density should favor rapid recombination, as is firmly predicted by PDR scenarios. At the same time, the oxygen remains in atomic form despite the significant extinction and high density that would tend to transform \element[0][]{O} to \element[][]{C}\element[][]{O}, depleting all the available carbon and also forming \element[][]{O}\element[][]{H} or \element[][]{H}$_2$\element[][]{O}. \par

\subsubsection{Chemical composition}
\label{comparison_with_other_scenarios}

Assuming that carbon might be only partially ionized, with a part as neutral atomic carbon or in a molecular form such as CO, would increase the amount of material not visible in any other tracer. In parallel, additional oxygen would have to be present in molecular form, presumable water ice, frozen out on dust. The attractive scenario of putting the extra carbon and oxygen into carbon monoxide can be ruled out observationally, as the low-J CO lines, in particular, those from rare isotopologues do not show a comparable column of cold CO. This analysis holds for the positions in M17~SW \citep{2015A&A...583A.107P}, where the \element[0][]{O} and \element[+][]{C} column densities are consistent with the elemental abundances. \par

For the positions where the derived column density ratio of oxygen to carbon is higher than the elemental abundance (i.e., positions 3 and 5 in M17~SW and both positions in Mon~R2), we can assume that carbon in the foreground absorbing material is indeed only partially ionized, with a fractional ionization of 20 to 50\%. Alternatively, the absorbing layer may be composed of two chemically different components, where a column density of about 20 to 50\% of the total would have to be low $T_\mathrm{ex}$, fully ionized \Cp{} material. For the Mon~R2 lines of sight, the assumption of additional absorbing material is perfectly reasonable. It is known that Mon~R2 is an embedded source with several young stellar objects \citep[YSO,][]{1976ApJ...208..390B} inside the molecular cloud illuminated from the backside. This darkening could help to explain the large discrepancy between the \element[0][]{O} and \element[+][]{C} column densities, compared to M17~SW, but does not explain the correlation between the \Oone~and \Cp{} emission, because we would not expect the presence of ionized carbon in the dense molecular foreground material. \citet{2012A&A...544A.110P} proposed in their model for Mon~R2, apart from the dense PDR, a surrounding envelope with densities of $5\times$ 10$^4$ cm$^{-3}$, molecular hydrogen column density of $5\times$ 10$^{22}$ cm$^{-2}$ and kinetic temperature of 35~K, similar values that we find here for the foreground layer and its estimated density from the differences between positions, confirming our results. Hence, we have for Mon~R2 a foreground separated in two phases, with a temperature gradient, where one part of the gas contains only \element[0][]{O} coexisting with molecular material. In contrast, in the other, outer layer \element[0][]{O} and \element[+][]{C} coexist. \par

M17~SW shows a different picture. The foreground column density ratios between \Oone~and \Cp{} are lower than the elemental ratio inside the main ridge and higher outside, showing some spatial correlation playing a role. As discussed in Sect~\ref{forebackcoldens}, studies over the whole map are needed. Single pointings are insufficient for a proper analysis of spatial effects. Moreover, \citet{2008ApJ...686..310H} found an optical foreground extinction of 2~mag, a significantly lower value than the foreground layer derived here. Therefore, we propose a scenario of a dense atomic foreground layer where ionized carbon and oxygen coexist with a density of 10$^3$ cm$^{-3}$, not ruling out the existence of a density gradient towards the exterior, leading to an additional atomic diffuse layer.

\subsection{Comparison between gas and dust}
\label{gasanddust}

We compare the column densities derived above from the \Oone~and \Cp{} fine-structure lines for the background and foreground components with the total gas column densities derived from the sub-mm-wavelength dust emission. The column density maps were derived from Herschel/PACS and SPIRE \citep{2010A&A...518L...3G} 250, 350, and 500 \um~ for Mon~R2 \citep{2017A&A...607A..22R} and M17~SW (Schneider, N., priv. comm.), using the method described by \citet{2013A&A...550A..38P}. The method estimates the total gas column density from the optically thin dust emission for both sources (Table~\ref{table:dust}) by fitting the spectral energy distributions (SEDs) and computes the gas surface density distribution of the region pixel by pixel. Note that this method deals with only a single temperature component of the dust emission and is thus biased towards the bulk column density of the lower temperature gas of the molecular clouds. It also traces parts of the warm PDR material, which has bright dust emission, but the analysis weighted it down by constraining the dust emission wavelength bands to longward of 160 \um. Also, the values estimated are upper values in a 18\arcsec~beam because there is line-of-sight contamination, with A$_{\mathrm{V}}\sim$10~mag for M17 and A$_{\mathrm{V}}\sim$2~mag for Mon~R2 (Schneider et al., in prep.).\par
 
For M17~SW, the gas column densities are between 54 and 336~magnitudes. Background equivalent extinction values derived from \Oone~at all positions are beween one and ten times smaller than the ones derived from dust. This difference is unsurprising as the dust emission also traces the dense molecular material of the bulk molecular cloud. The warm PDR traced by the background emission in the fine structure lines contributes 10 to 25\% of the total column density traced by the dust emission (Table~\ref{table:dust}). We also note that the column density of the warm PDR background derived here is larger than the one derived from the mid-and high-J CO lines \citep{1987ApJ...322L..49H,2015A&A...583A.107P} and quoted in previous studies of the \Cp{} emission \citep{1988ApJ...332..379S}. The low-velocity resolution of the last \Cp{} observations ignored that the foreground absorption significantly weakens the \Cp{12} emission. We note that the cold and dense molecular cloud material responsible for the bulk dust emission is not traced by the column densities derived from the \Oone~lines observed in emission. This disparity is plausible as oxygen will be mainly in molecular form (\element{C}\element{O}, \element{H}$_2$\element{O}) and partially frozen out onto dust grains. \par

\begin{table*}
\centering
\caption{Mon~R2 and M17~SW dust equivalent extinction comparison.}
 \label{table:dust}
 \begin{threeparttable}
\begin{tabular}{c | c | c c c c | c c c c }
\hline
\hline
 &   & \multicolumn{4}{c|}{Background} & \multicolumn{4}{c}{Foreground} \\
 \hline
Component & A$_{\mathrm{V}}$ Dust & A$_{\mathrm{V,bg}}$\element[0][]{O} \tnote{a}& $\frac{\mathrm{A}_{\mathrm{v}} \mathrm{Dust}}{\mathrm{A}_{\mathrm{v,bg}}\element[0][]{O}}$ & A$_{\mathrm{V,bg}}$\element[+][]{C}\tnote{a} & $\frac{\mathrm{A}_{\mathrm{v}} \mathrm{Dust}}{\mathrm{A}_{\mathrm{v,bg}}\element[+][]{C}}$ & A$_{\mathrm{V,fg}}$\element[0][]{O}\tnote{b} & $\frac{\mathrm{A}_{\mathrm{v}}\mathrm{Dust}}{\mathrm{A}_{\mathrm{v,fg}}\element[0][]{O}}$ & A$_{\mathrm{V,fg}}$\element[+][]{C}\tnote{b} & $\frac{\mathrm{A}_{\mathrm{v}}\mathrm{Dust}}{\mathrm{A}_{\mathrm{v,bg}}\element[+][]{C}}$    \\
  & (mag.) & (mag.) & & (mag.) & & (mag.) & & (mag.) &  \\
         \hline
MonR2 1 & 73 & 106 & 0.7 & 19 & 3.8 & 9.7 & 7.5 & 3.7 & 20 \\      
MonR2 2 & 40 & 28 & 1.4 & 21 & 2.9 & 8.1 & 4.9 & 2.9 & 14 \\
       \hline
M17SW 0 & 253 & 28 & 9.0 & 41 & 12 & 3.8 & 64 & 9.2 & 28  \\
M17SW 1 & 336 & 105 & 3.2 & 36 & 9.3 & 6.8 & 49 & 7.6 & 44 \\
M17SW 2 & 282 & 82 & 3.4 & 25 & 11.3 & 7.1 & 40 & 7.6 & 44 \\
M17SW 3 & 328 & 28 & 12 & 20 & 16  & 6.2 & 53 & 3.5 & 94 \\
M17SW 4 & 167 & 20 & 8.4 & 34 & 4.9 & 5.1 & 33 & 5.8 & 29  \\
M17SW 5 & 54 & 27 & 2.0 & 13 & 4.2 & 8.2 & 6.6 & 1.7 & 32 \\
M17SW 6 & 88 & 31 & 2.8 & 34 & 2.6 & 5.3 & 17 & 8.0 &  11 \\
        \hline
\end{tabular}
\begin{tablenotes}\footnotesize
\item[a] Extracted from Table~\ref{table:OICIItable_back}.
\item[b] Extracted from Table~\ref{table:OICIItable_fore}.
\end{tablenotes}
\end{threeparttable}
\end{table*}

For Mon~R2, the situation is different. The equivalent extinctions derived from dust for both positions are much closer to those derived from the \Oone~fine structure lines. This closeness could be interpreted as neutral oxygen that coexists with dense foreground molecular material, as has been suggested before by several authors \citep{2012A&A...544A.110P,2012A&A...543A..27G,2019A&A...629A..81T}. As discussed above in Sect~\ref{foreground_physical}, the dust strengthens the scenario of a double-layered foreground, with a molecular component and another in atomic form. \par

The comparison of the column densities derived for the emission from the warm background in \Cp{} and \Oone~shows that both are consistent with the emission originating in standard PDR material, with excitation temperatures between 60 and 140~K for the \Oone~and between 150 and 250~K for the \Cp{}. As discussed in our previous analysis, the amount of material is significantly larger than expected for a single PDR layer, so we have to invoke multiple PDR layers within the beam with several tens of magnitudes in $A_v$. The column densities derived here refer only to the bright PDR material in the fine structure line emission. A portion of the warm, dense PDR material will contribute to the mid- and high-J CO emissions. These lines can trace the material because we have worked under the assumption that all \Oone~ emission is in the form of atomic material, not considering the molecular contribution.  \par 

\subsection{Comparison with other models}
\label{comparison_models}

At his point, we want to consider whether there are alternative explanations for the nature of the foreground layer. \citet{2021ApJ...916....6G} discussed the \Oone~profile toward W3, which shows strong absorption, similar to what this paper discusses. At first sight, the situation looks similar to the Mon~R2 case discussed here. They found similar values for the neutral oxygen foreground column density, from 2 to 7$\times$10$^{18}$~cm$^{-2}$, reinforcing the idea of a foreground layer of atomic material with similar column densities. They attribute its foreground absorption to geometry effects because the source is heated from the backside, allowing for additional oxygen in atomic form in cold molecular layers, together with CO, between the PDR and the observer. However, such a scenario cannot explain the sizeable \element[+][]{C} column densities together with more prominent \element[0][]{O} column densities at the same velocity and line width. The \Cp{} absorption dips, requiring a large column of \Cp{} much colder than feasible in a PDR layer, rule out a scenario similar to the one discussed for W3 by \citet{2021ApJ...916....6G}. The analysis continues in \citet{2023ApJ...952..102G}, with the modeling, through the Meudon code \citep{2006ApJS..164..506L}, of the foreground layer in W3A. Their model identifies foreground hydrogen column densities of 2$\times$10$^{22}$~cm$^{-2}$, or equivalently a oxygen column density of $\sim$ 5$\times$10$^{22}$~cm$^{-2}$, with a density of n(\element[][]{H}) = 250 cm$^{-3}$ over a spatial extension of 26~pc. While the interpretation of the absorption profile in \Oone~resulting from a smooth foreground may be a valid scenario in the case of W3A, the \Oone~absorption observed in Mon~R2 and M17~SW must be due to high-density material; this is based on the differences in foreground column densities between positions over small angular distances. Mapping observations in \ Oone~ are not available for W3A; hence, a decision between these two scenarios is impossible with the presently available data. In addition, a \Cp{13} profile with sufficient S/N is not available in the case of W3A, so the background column density of \Cp{} is not well constrained. W3A might behave like our sources, but maps in \Oone~and \Cp{} would be needed to distinguish between the scenarios. Regretfully, new observations in these atomic cooling lines will not be feasible in the near future. \par

\citet{2022A&A...659A..36K} presented a similar scenario for RCW120 for \Cp{} only. \Cp{} observations show correlations between the velocity of foreground components derived from a similar multi-component double-layered Gaussian analysis and atomic hydrogen absorption dips in the line profile. Through a \ion{H}{I} self-absorption (HISA) analysis, they compared the \ion{H}{I} absorption line profile with that of \Cp{}. They found correlations in line width and LSR velocity for some components. They attributed the origin of the foreground layer causing the \Cp{} self-absorption to diffuse and extended \ion{H}{I} that coexists with \element[+][]{C}. Comparison of the foreground column density between adjacent positions and at an assumed temperature of 15~K, similar to the 20~K we use (Table 1 of \citet{2022A&A...659A..36K}), show a column density difference of 0.5-1$\times$10$^{21}$~cm$^{-2}$, much lower than the values we derive for M71~SW and Mon~R2 in Sect.~\ref{denstiy_estimate}. Similarly, in the case of RCW120, the difference in the foreground line profile parameters between positions (Fig.~8 of the paper as mentioned earlier),  namely the intensity and velocity position and width, are much smaller for RCW120, compared to the case of M17~SW and Mon~R2, showing a much more homogeneous foreground than for our sources. \OI{63}~was observed in the source (Kabanovic, S., priv. comm.), but there is only faint emission in the southwestern area. Besides, no \OI{145} observations are available, making verifying any self-absorption effects in the \OI{63}~profiles difficult. New observations in high-resolution S/N for oxygen lines would be needed for an analysis similar to ours.\par

\section{Summary} \label{Summary}

The newly observed high spectral resolution observations of the \OI{145} and the \OI{63} lines towards M17~SW and Mon~R2 show strong absorption notches in the \OI{63} line and allow for a detailed multi-component Gaussian analysis, separating the emission into a background and foreground layer. The background emission is consistent with warm, dense PDR-material emission in both sources, which is well-known and characterized by many previous studies. This material shows smooth Gaussian emission profiles towards which we formerly observed intense \Cp{12} self-absorption in both sources and now newly detected \OI{63} foreground absorption. The deep foreground self-absorption features require significant column densities of low $T_\mathrm{ex}$ atomic oxygen in the foreground material.\par

Comparison of the \Oone~profiles with the previously observed \Cp{12} and \Cp{13} profiles towards both Mon~R2 and M17~SW show that the \OI{145} emission line profiles, tracing the warm background, are entirely consistent with the \Cp{13} emission profiles from the background material and have similar total column densities. The intense self-absorption notches visible in \OI{63} match well with the previously observed \Cp{12} absorption features, both in their central velocities and widths. Hence, we conclude that (for the majority) they trace the same material, which has to be very cold (in terms of excitation temperature) due to the deep absorption. The derived column densities in the absorption layer are also consistent for both species. The physical parameters of the background layer, though the column density derived can vary by a factor of up to 10 depending on the detailed assumptions, are, for both species, entirely consistent with a multi-layer PDR at the UV intensity and density of the bulk material in M71~SW and Mon~R2.\par

The physical parameters of the foreground layer are well-constrained based on the analysis presented. The large column density of the foreground material and the small spatial scales derived from the sudden variation of the central velocity and depth of the absorption features constrain the foreground material to have a significant density between $10^3$ and $10^4\,{\rm cm^{-3}}$. The nature of this foreground layer, with its fully ionized carbon and atomic oxygen at relatively high densities and low excitation temperature, is very puzzling. \par

We emphasize that the foreground absorption is only visible at the high spectral and spatial resolution available with the upGREAT instrument on SOFIA and Galactic sources. Spatial averaging, namely, observations of sources at a larger distance, namely in nearby or further out galaxies, resulting in lower spectral resolution observations, smooth out the absorption features in the line profiles, leading to reduced integrated intensity. Thus, the absorbing foreground absorption component of unidentified origin may be partially responsible for the \Oone~and \Cp{} line intensity deficiencies quoted for such sources. Large-scale mapping observations with high signal-to-noise in both \Cp{} and \Oone~fine structure lines and their analysis in galactic clouds are necessary to resolve this issue. We analyzed one of the best datasets that combines both \Cp{} isotopes with both \Oone~transitions, and without access to new observations at the exact resolution and S/N in the near future, it will be hard to disentangle this scenario. Still, we can affirm without doubt that there is a cold, relatively dense foreground layer associated with the main PDR composed of ionized carbon and neutral oxygen absorbing the emission of the background.
                
\begin{acknowledgements}
This work is based on observations made with the NASA/DLR Stratospheric Observatory for Infrared Astronomy (SOFIA). SOFIA was jointly operated by the Universities Space Research Association, Inc. (USRA), under NASA contract NNA17BF53C, and the Deutsches SOFIA Institut (DSI) under DLR contract 50 OK 0901 to the University of Stuttgart. The work was carried out within the Collaborative Research Centre (CRC) 956 sub-projects A4 and C1 (project ID 184018867) and CRC 1601 sub-projects A6 and B2, funded by the Deutsche Forschungsgemeinschaft (DFG), project ID 500700252. This research was carried out in part at the Jet Propulsion Laboratory, California Institute of Technology, under contract with the National Aeronautics and Space Administration (80NM0018D0004). S.K. acknowledges support from the Orion-Legacy project that is supported by the BMWI via DLR, project number 50OR2311. 
\end{acknowledgements}

\bibliographystyle{aa} 
\bibliography{bibpap_OI} 

\begin{appendix}

\section{Observational parameters}\label{app:dpeac}

In this appendix, we provide the observational parameters of the observed positions in detail. The coordinates of each position are the same as the original \Cp{} observations from \citet{2020A&A...636A..16G}. The tables contain the absolute and relative coordinates (with respect to the source coordinate, see Table~\ref{table:tsys}) for each observed position and their OFF position.

\begin{table}[ht]
  \centering
    \caption{Monoceros~R2 and M17~SW positions.}
     \begin{threeparttable}
  \begin{tabular}{l r r r r  }
      \hline
      \hline
      & RA      & DEC                     & Rel.       & Rel.        \\
            &  &                       & Offset     & Offset  \\
            & (J2000) & (J2000) & RA & DEC\\
            & (h:m:s) & (\degr:':'') &  ('') &  ('')      \\
            \hline
Mon~R2 & \\
Pos. 1  & 6:07:46.21 & $-$6:23:03.01 & 0 & 5        \\
Pos. 2  & 6:07:44.87 & $-$6:23:03.01 & -20 & 5      \\
OFF    & 6:07:19.28 & $-$6:33:08.72 &  -400 & -600 \\
\hline
M17~SW & \\
Pos. 0  & 18:20:23.46 & $-$16:12:02.01 & $-$60.6 & $-$1.1     \\
Pos. 1  & 18:20:22.45 & $-$16:12:30.20 & $-$75.1 & $-$29.3    \\
Pos. 2  & 18:20:21.34 & $-$16:12:05.00 & $-$91.1 & $-$4.1     \\
Pos. 3  & 18:20:22.32 & $-$16:11:35.89 & $-$76.9 & 25.0     \\
Pos. 4  & 18:20:24.61 & $-$16:11:34.82 & $-$44.0 & 26.1    \\
Pos. 5  & 18:20:25.70 & $-$16:12:02.77 & $-$28.2 & $-$1.9    \\
Pos. 6  & 18:20:24.66 & $-$16:12:29.23 & $-$43.3 & $-$28.3   \\
OFF    & 16:21:04.87 & $-$16:13:07.35 & 537     & -67 \\
        \hline
  \end{tabular}
\end{threeparttable}
\end{table}  

\section{Excitation temperature estimate for the \Oone~line for the background} \label{Appendix:tex}

We have estimated the excitation temperature for both \Oone~ transitions in a three step process. We do not have previous information about the excitation temperature of the oxygen; therefore, as a starting point we assume that \OI{63}~is optically thick so that the brightness temperature traces the excitation temperature (including the Rayleigh-Jeans correction), and at least at one velocity, the background brightness temperature shines through the foreground absorption. Hence, we have selected as a lower limit for the \OI{63} the T$_\mathrm{mb}$ peak emission and then converted to T$_\mathrm{ex}$ given by the Rayleigh-Jeans correction under the assumption of optically thick emission as:

\begin{equation}
T_\mathrm{mb} =  \mathcal{J}_{\nu}(T_\mathrm{ex}) = \frac{h\nu}{k}\left(e^{h\nu/kT_\mathrm{ex}}-1\right)^{-1}
\label{eq:rj}
.\end{equation}

With the excitation temperature of the \OI{63} line, we can estimate the excitation temperature of the \OI{145} line, though the balance between collisional excitation and de-excitation and spontaneous emission when assuming a gas density. Then, the collisional rate $C_{ij}$ from level $i$ to $j$ is given by:

\begin{equation}
C_{ij} = R_{ij} n
\label{eq:cij}
,\end{equation}

with $n$ the density of the collision partner (cm$^{-3}$) and $R_{ij}$ the collisional rate coefficient (cm$^{-3}$~s$^{-1}$¸). The oxygen coefficients at different kinetic temperatures for electrons, hydrogen, and molecular hydrogen were estimated by \citet{2018MNRAS.474.2313L}. Now, the ratio between adjacent levels for a three-level system is \citep{2015ApJ...814..133G,2019ApJ...887...54G}:

\begin{equation}
\frac{n_2}{n_1} = \frac{{C}_{12}\left({C}_{01}+{C}_{02}\right) + {C}_{02}\left({A}_{10}+{C}_{10}\right)}{\left({A}_{21}+{C}_{21}+{C}_{20}\right)\left({C}_{01}+{C}_{02}\right)-{C}_{20}{C}_{02}}
\label{eq:n2ton1}
,\end{equation}

\begin{equation}
\frac{n_1}{n_0} = \frac{\left({A}_{21}+{C}_{21}+{C}_{20}\right)\left({C}_{01}+{C}_{02}\right)-{C}_{20}{C}_{02}}{\left({A}_{21}+{C}_{21}+{C}_{20}\right)\left({A}_{10}+{C}_{10}\right)+{C}_{20}{C}_{12}}
\label{eq:n1ton0}
.\end{equation}

The excitation temperature of a transition is related to the ratio of the level population through the Boltzmann equation  by: 

\begin{equation}
\frac{n_j}{n_i} = \frac{g_j}{g_i} \exp\left[-\left(E_j - E_i \right)/k_{\mathrm{b}}T_{\mathrm{ex}ij}\right]
\label{eq:lp}
,\end{equation}

with $g$ the statistical weight and $E$ energy of a level. Upward and downward collision rates are related by:

\begin{equation}
C_{ji} = C_{ij} \frac{g_i}{g_j} \exp\left(-h \nu_{ij} /k_{\mathrm{b}} T_{\mathrm{kin}} \right)
\label{eq:crates}
.\end{equation}

\begin{figure}
   \centering
   \includegraphics[width=1\hsize]{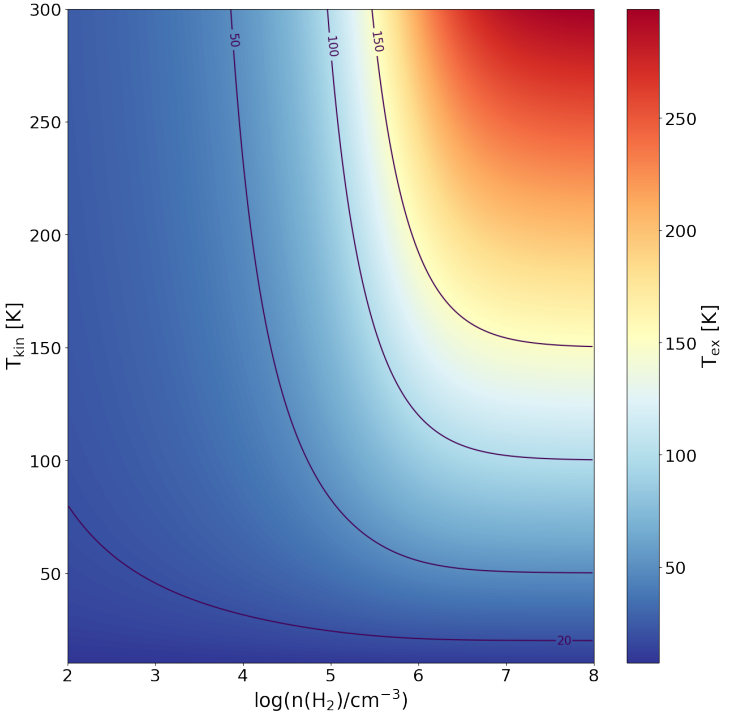}
      \caption{Excitation temperature of \OI{63} for a combination of kinetic temperature and density of the gas. We have added contour levels for excitation temperatures of 20, 50, 100 and 150~K. }
         \label{fig:63_tex}
\end{figure}

\begin{figure}
   \centering
   \includegraphics[width=1\hsize]{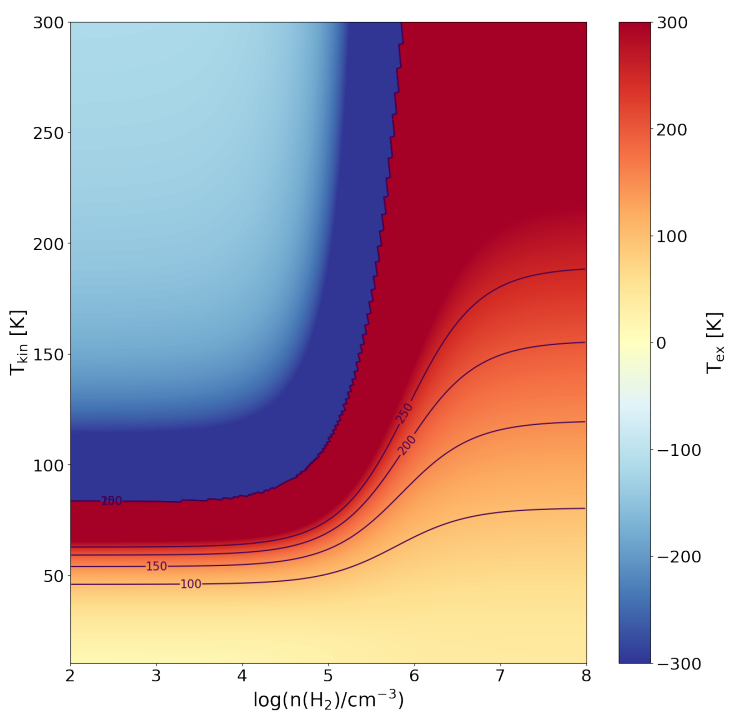}
      \caption{Excitation temperature of \OI{145} for a combination of kinetic temperature and density of the gas. We have added contour levels for excitation temperatures of 100, 150, 200 and 250~K.}
         \label{fig:145_tex}
   \end{figure}

Here, we selected molecular hydrogen as main collision partner with an orto-to-para ratio of 3:1. The resulting excitation temperatures for \OI{63} and \OI{145} as a function of kinetic temperature and density are shown in Fig.~\ref{fig:63_tex} and \ref{fig:145_tex}, respectively. 

\begin{figure}
   \centering
   \includegraphics[width=1\hsize]{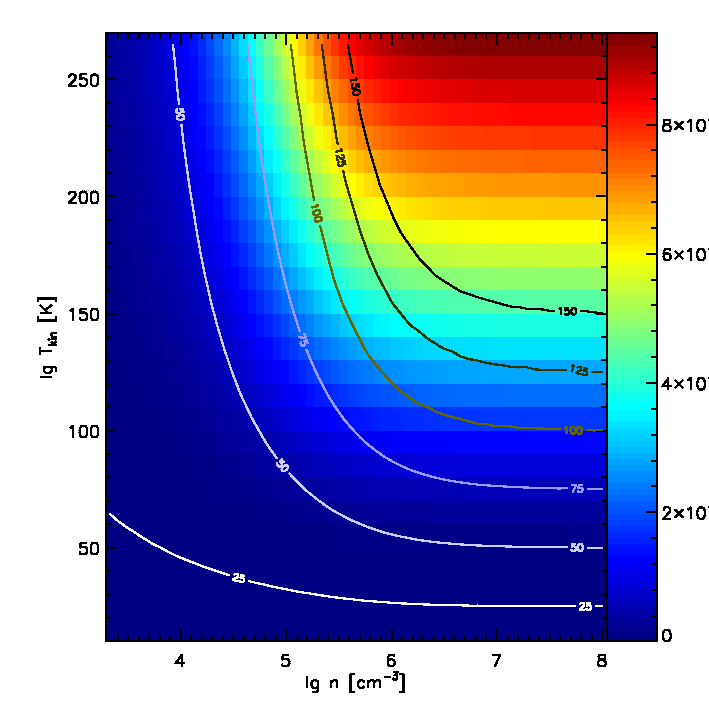}
      \caption{\OI{145} emissivity in colors for a given column density. The contours are the \OI{63} excitation temperature as in Fig.~\ref{fig:63_tex}.     
      }
         \label{fig:oi_sensitivity}
   \end{figure}

Using these dependencies we can obtain the \OI{145}~excitation temperature from the \OI{63}~excitation temperature if we assume either a particular gas density or kinetic temperature. We can simply look up the parameters in Fig.~\ref{fig:63_tex} that produce the \OI{63} excitation temperature measured in the first step and go with these parameters into Fig.~\ref{fig:145_tex} to obtain the \OI{145} excitation temperature. Here, we assumed a density of 10$^6$ cm$^{-3}$, as an intermediate value given the critical density of both transitions. A different method would be to fix the kinetic temperature and then search for a density. The methods are equivalent and the actual choice of  the method or the value are not significant when it comes to the final output. Actually, it is not necessary to have a precise value for the \OI{145}~excitation temperature because the quantity entering the radiative transfer is the emissivity ($B(T_\mathrm{ex}$) $\times$ $\tau_{145}$) -- and not $T_\mathrm{ex}$ as such. In Fig.~\ref{fig:oi_sensitivity}, we plot in colors the \OI{145}~emissivity ($B(T_\mathrm{ex}$) $\times$ $\tau_{145}$) as a function of the kinetic temperature and column density on a Rayleigh-Jeans scale. We can see that the \OI{145}~emissivity and the \OI{63} excitation temperature go almost in parallel. Thus, if we have a given $T_{\mathrm{ex},63}$, we can read the  \OI{145}~emissivity with reasonable accuracy. We do not need to know the exact excitation temperature of the \OI{145} line and the corresponding gas density and temperature. We see that for a $T_{\mathrm{ex},63}$ of 75~K and below, the \OI{145}~emissivity stays almost constant along the contours, it is independent of the actual density. Hence, we have no significant uncertainty in the background column from the method. At higher $T_{\mathrm{ex},63}$ we can conclude that we are still safe against variations towards higher densities, but that, for example, for $T_{\mathrm{ex},63}$ = 100~K, we underestimate the emissivity by a factor 2 if the actual gas density is 10$^5$ cm$^{-3}$ instead of 10$^6$ cm$^{-3}$ so that we overestimate the background component column by this factor of two. \par

Altogether, we have estimated the excitation temperature in a 3-step process: 1) selection of an excitation temperature from \OI{63} main beam temperature peak, 2) estimation of the kinetic temperature of oxygen from Fig.~\ref{fig:63_tex} under the assumption of a density of 10$^6$ cm$^{-3}$, together with the excitation temperature from 1), 3) estimate of the excitation temperature for \OI{145} from Fig.~\ref{fig:145_tex}. The derived excitation temperatures for both transitions and the kinetic ones are shown in Table~\ref{table:Tex_Tkin}.

\section{Variation of parameters for background and foreground}

\subsection{Effects of varying the density of the background layer} \label{app:densityeff}

\begin{figure}
   \centering
   \includegraphics[width=1.\hsize]{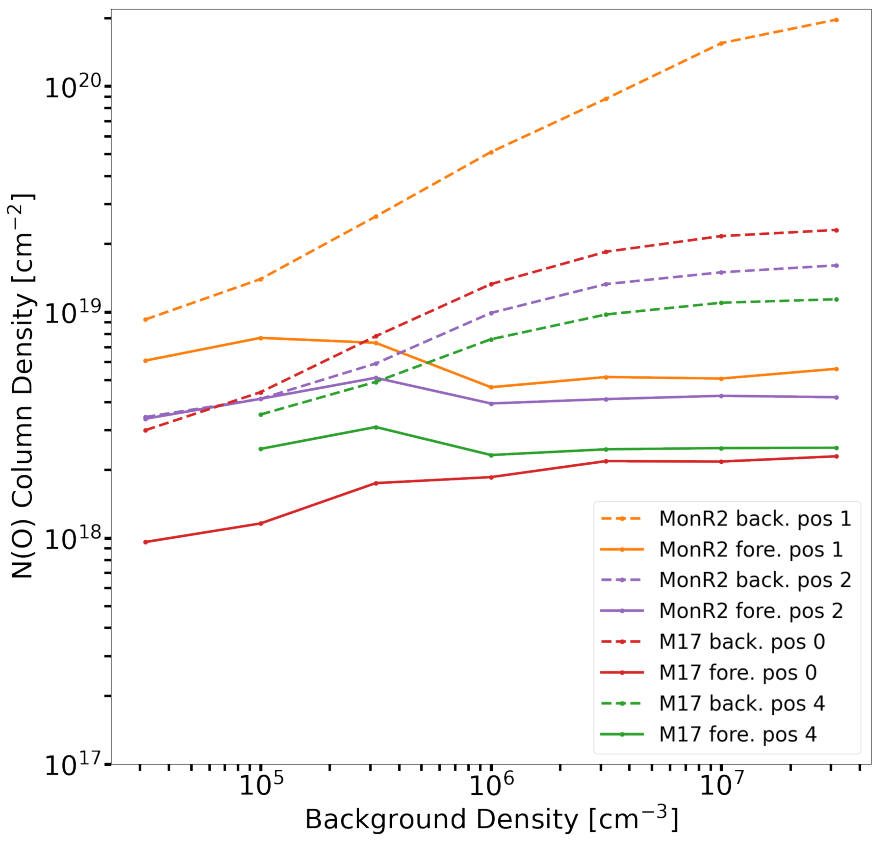}
      \caption{Best-fit column densities compared for selected positions against a range in the assumed density. The background is indicated by dashed lines and the foreground in solid ones.
              }
         \label{fig:allTexN}
   \end{figure}

In this subsection, we analyze the dependency of the derived physical parameters on the assumed value for the density of the background layer. As the default, we have considered a density of 10$^6$ cm$^{-3}$ for deriving the upper transition excitation temperature from the lower transition excitation temperature, estimated from the peak brightness of the \OI{63} line. Figure~\ref{fig:allTexN} shows how changes in the assumed collision partner density of the background, and hence different values for the derived $T_{\mathrm{ex},145}$ (derived from the analytic solution of the three-level system in Appendix~\ref{Appendix:tex}), result in different best-fit results for the background (dashed line) and foreground column density (solid line). We have selected two positions for each source to be analyzed in exploring the boundaries of our solutions. The other positions are affected in the same way.

The analysis shows that an increase in density leads to a corresponding increase in the estimated background column density and vice versa. This proportionality can be explained according to Fig.~\ref{fig:145_tex}, which illustrates the effect that the upper \Oone~line shows a regime of population inversion. Any increase in density results in a decrease in the \OI{145} excitation temperature (and vice versa) and, consequently, an increase in the \Oone~column density necessary to maintain the observed intensity. In comparison, the \OI{63} excitation temperature remains constant. Note that M17~SW position 4 can only be fitted with densities higher than 10$^5$~cm$^{-3}$. For lower densities, the \OI{145}~peak intensity cannot be reached. It is worth noting that position 4 has the highest \OI{145} intensity and is located at the \Oone~peak, suggesting that the density here should be much higher than the other positions. \par 

The foreground column density necessary for the deep absorption features depends very little on the background layer's density variation, as is illustrated in Fig~\ref{fig:allTexN}. This insensibility is evident as the \OI{63} absorption is highly optically thick so that the increase in the background intensity to be absorbed does not need to be compensated by an enhanced foreground absorption. The modest variations visible in the fit result originate in second-order effects. Depending on the details of the line profiles, the optically thinner wings of the absorbing layers are more or less sensitive to an increase or decrease in the background intensity resulting from the density variations.

\subsection{Effects of varying the kinetic temperature of the background layer}
\label{app:kinetic}

Now, let us continue analyzing the sensitivity of the derived background and foreground column densities against the second physical parameter, the kinetic temperature of the background, keeping the density fixed at $10^6\rm{cm}^{-3}$. Above, we have assumed the kinetic temperature derived from the upward Rayleigh-Jeans corrected peak brightness temperature of the \OI{63}~emission shining through between the absorption dips, giving a lower limit to the excitation temperature. As the kinetic temperature is always larger than the excitation temperature (see Fig.~\ref{fig:145_tex}) for any density, this is also a lower limit to the kinetic temperature of the background. We now vary the kinetic temperature of the background from this minimum value up to higher values. We calculate the excitation temperatures for both \Oone~lines from the analytic formulas at the given temperature and density (see Appendix~\ref{Appendix:tex}) and perform the least-squares fit of the line profiles with these parameters. \par

We stop increasing the kinetic temperature of the background once the background column density drops to the level of the foreground, breaking our initial assumption of having most of the material concentrated in the warm background layer. This upper limit gives the maximum kinetic temperature used in the fitting, roughly 50~K above our minimum value. Therefore, we have a reduced range of about 50~K for valid kinetic temperatures where our assumptions remain accurate. \par

\begin{figure}
   \centering
   \includegraphics[width=1.\hsize]{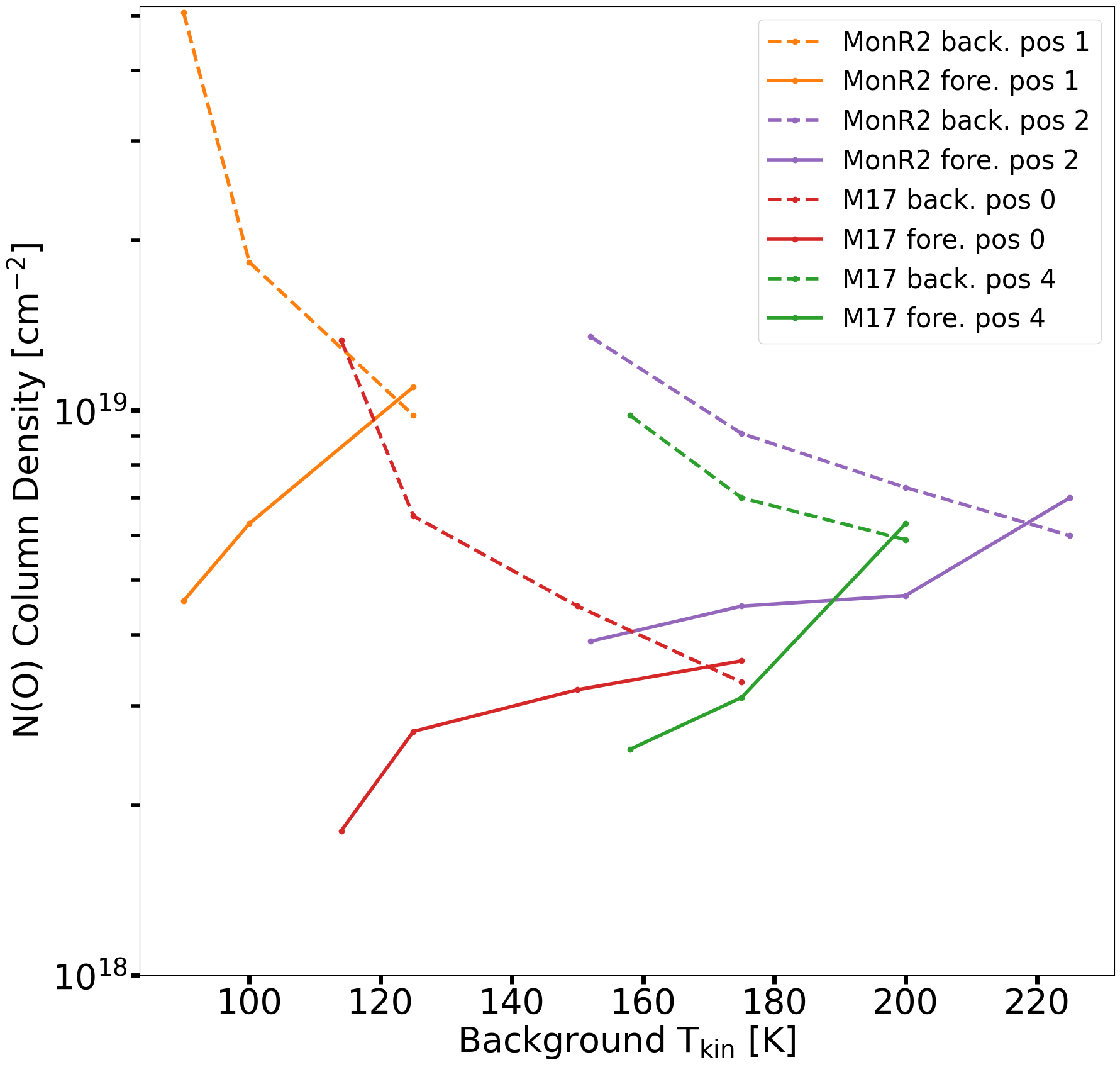}
      \caption{Best fit column densities compared for selected positions against a range in the assumed kinetic temperatures. 
              }
         \label{fig:allTex}
   \end{figure}

Figure~\ref{fig:allTex} shows the thus fitted column densities for both layers against the range of background kinetic temperatures for selected positions fitted in Mon~R2 and M17~SW. The figure shows that to fit the observed brightness of the background emission; the necessary background column density is roughly inversely proportional to the excitation temperature and, hence, the kinetic temperature.

The foreground column density is only slightly affected by the changes in the background. Increases in the background kinetic temperature lead to a rise in the background excitation and main beam temperatures, reducing the optical depth due to the severe reduction in column density. Then, the higher background intensity requires more foreground material to absorb the background emission and obtain the observed self-absorbed profile. In any case, the range of variation is lower than an order of magnitude, keeping the foreground density well-constrained.\par

\subsection{Effect of variations in the foreground excitation temperature} \label{app:foresect}

\begin{figure}
   \centering
   \includegraphics[width=1.\hsize]{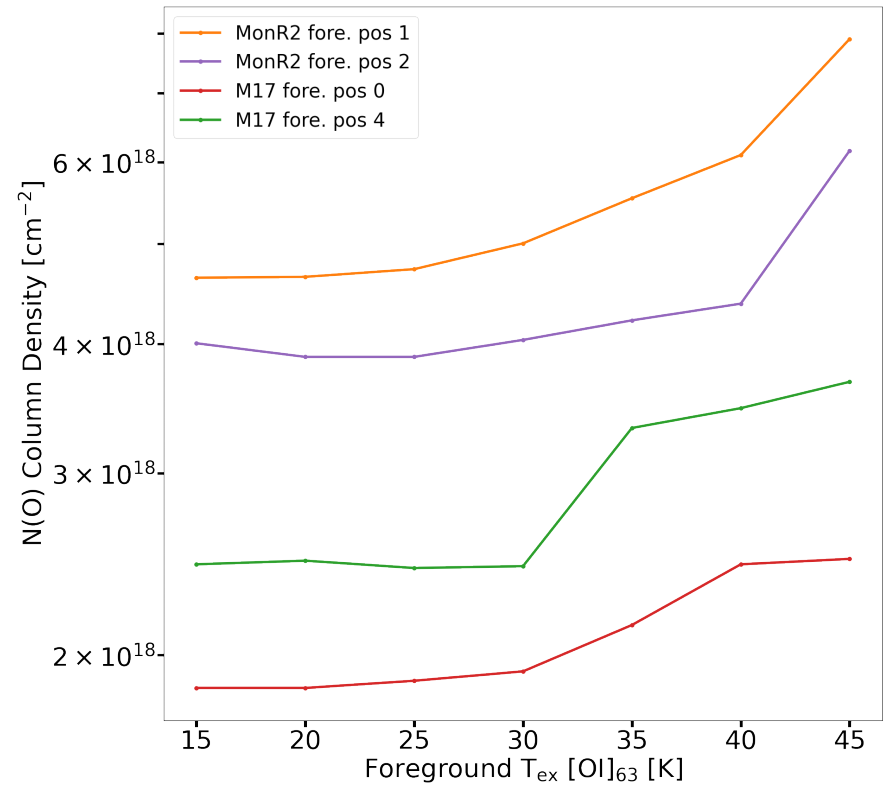}
      \caption{Best fit foreground column densities compared for selected positions against a range in the assumed foreground excitation temperatures. 
              }
         \label{fig:allTexf}
   \end{figure}

We now analyze how changes in the foreground excitation temperature assumed for the \OI{63} line affect the foreground column density. For the previous analysis, we have assumed a temperature of 20~K. Figure~\ref{fig:allTexf} shows how the foreground column density derived from the line profile changes when the foreground's excitation temperature varies. We varied the foreground excitation temperature from 15~K to 45~K and selected 15~K as a lower limit because this value is the lowest possible for \Cp{} from the energy balance \citep{2022A&A...659A..36K}, a reasonable assumption where both atoms should coexist. There is insufficient material to absorb the background in some positions for lower values than this, establishing a lower limit to the foreground excitation temperature. The upper value of 45~K is an appropriate upper limit for the foreground gas to produce visible absorption nodges still; higher values come too close to the background values. \par  

\section{Parameters of individual components}\label{app:tables}

\begin{figure}
   \centering
\includegraphics[width=1\hsize]{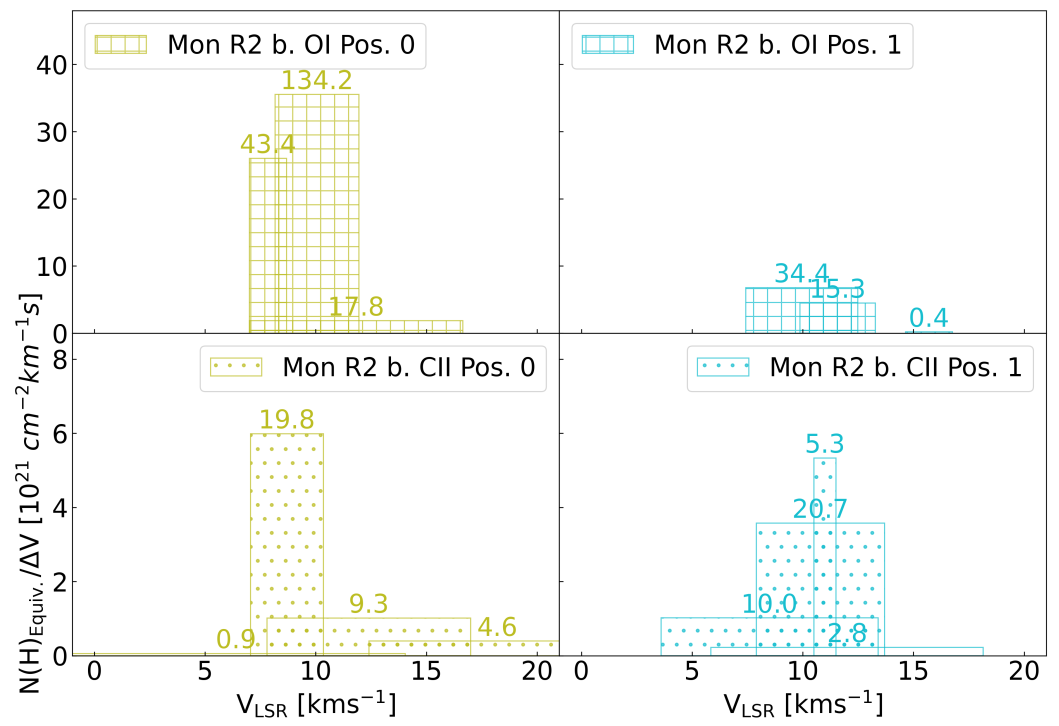}
\caption{The fitted components of the background emission in Mon~R2 at each position observed with high S/N, both for \Oone~(this paper; top) and \Cp{} (from the former study by \citet{2020A&A...636A..16G}, bottom) at different scales. The components are shown as bars with an amplitude given in column density per velocity width on an equivalent $N_{\ion{H}{I}}$ column density scale. The width of each bar is the FWHM width of the component. The (integrated) column density of each component is indicated by the number at the top.}
\label{barplot_bg_MonR2}
\end{figure}

\begin{figure}
   \centering
\includegraphics[width=1\hsize]{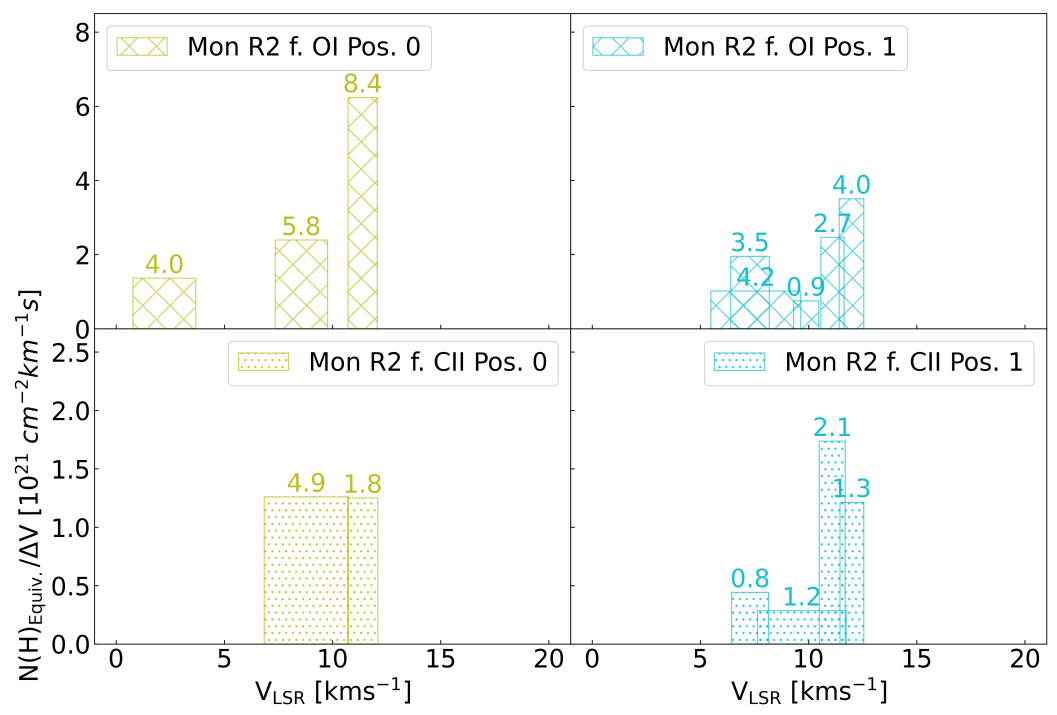}
\caption{Same as Fig.~\ref{barplot_bg_MonR2}, but for the foreground components at the two positions observed for Mon-R2.}
\label{barplot_fg_MonR2}
\end{figure}

\begin{figure*}
   \centering
\includegraphics[width=1\hsize]{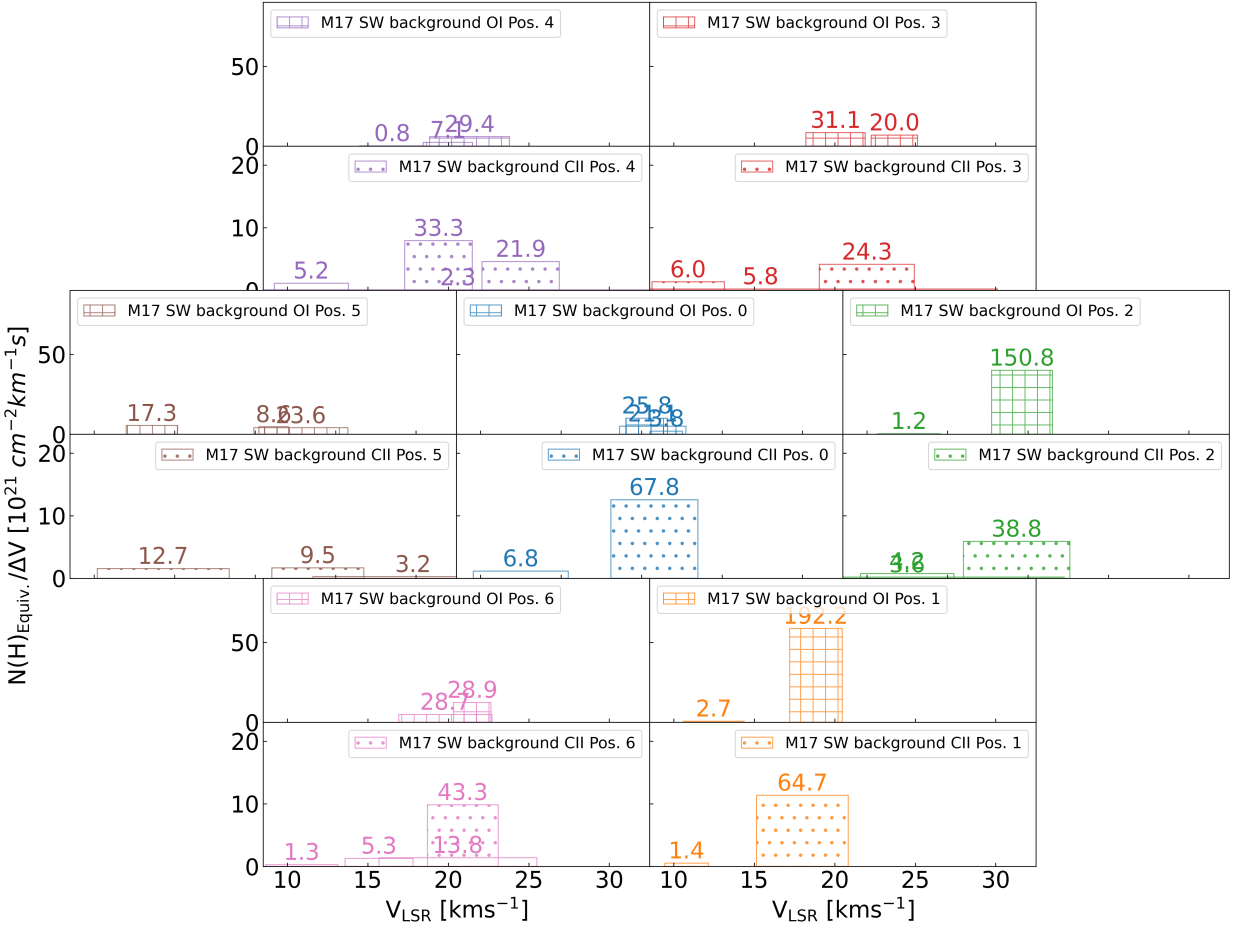}
\caption{Same as Fig.~\ref{barplot_bg_MonR2}, but for the background components fitted at the M17~SW positions.}
\label{barplot_bg_M17}
\end{figure*}

\begin{figure*}
   \centering
\includegraphics[width=1\hsize]{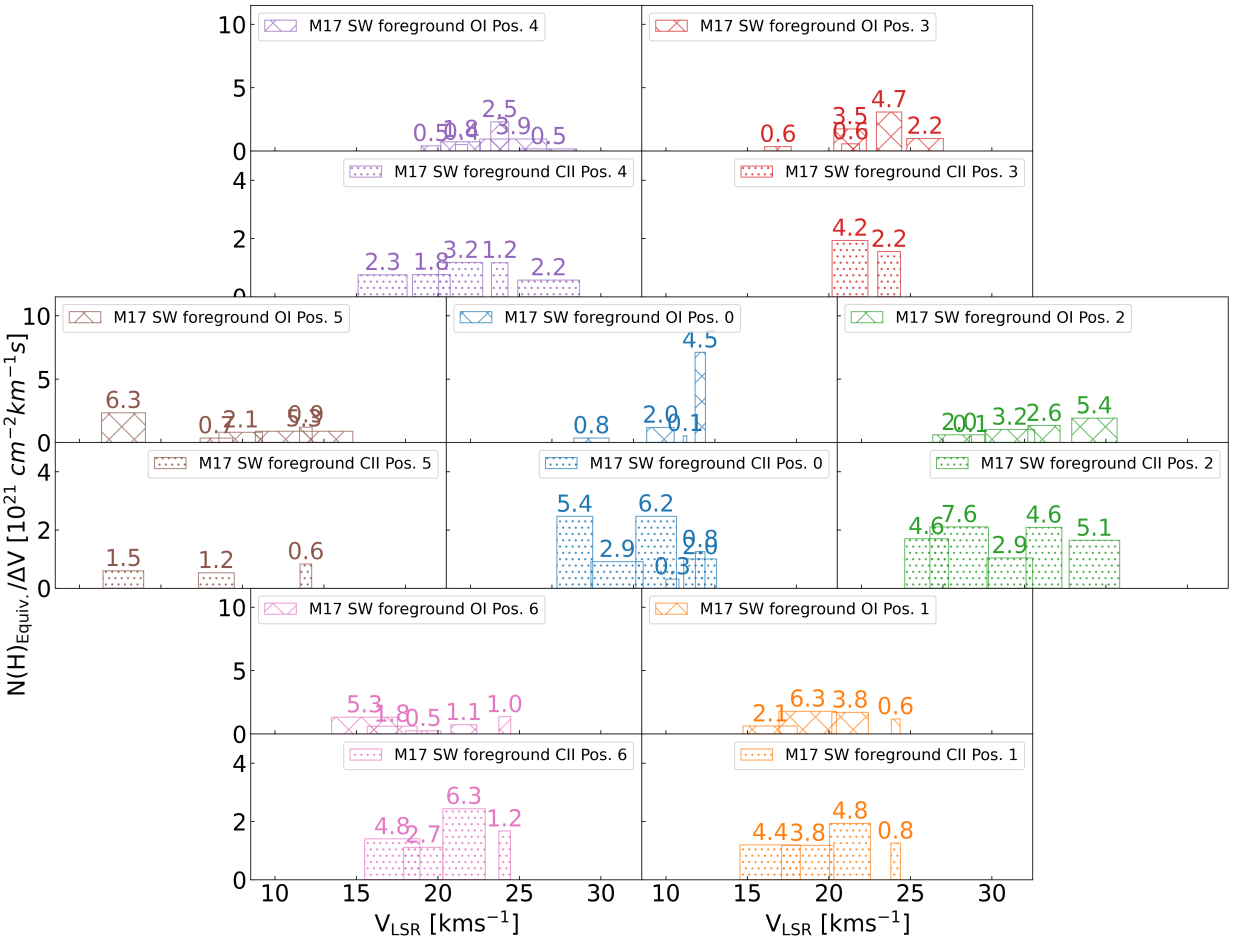}
\caption{Same as Fig.~\ref{barplot_bg_M17}, but for the foreground components fitted at the M17~SW positions.}
\label{barplot_fg_M17}
\end{figure*}

In this appendix, we list the fitted parameters for each Gaussian component for all the positions according to equation~\ref{eq:tmb}, \ref{eq:tauNu4} and \ref{eq:phi2}. Tables~\ref{app:monr2_1} and \ref{app:monr2_1a} describe Mon~R2 and Tables~\ref{app:m17_1} to \ref{app:m17_7} describe M17~SW. Tables~\ref{app:monr2_1_fixed1} to \ref{app:m17_6_fixed2} describe the attempts of using foreground parameters of one position into another. The main components, providing more than 90\% of the background and foreground material are indicated by numbers in bold print. 
\textbf{For better visualization of the fit results, we present the fitted parameters for each of the Gaussian components for each source position in a bar plot in Figures~\ref{barplot_bg_MonR2} to \ref{barplot_fg_M17}.}

\begin{table*}[ht]
\centering
\caption{Mon~R2 position 1 parameters.}
 \begin{threeparttable}
\begin{tabular}{c c c c | c | c }
\hline
\hline
Component & N$_u$ \Oone & V & $\Delta$V & N$_l$ \Oone\tnote{a} & N \Cp{}\tnote{b} \\
 & (cm$^{-2}$) & (km~s$^{-1}$¸) & (km~s$^{-1}$¸) & (cm$^{-2}$) & (cm$^{-2}$)  \\
\hline
\multicolumn{4}{l}{T$_{\mathrm{ex}}$ Background [OI]$_{145}$ = 144.6 K} \\
Comp. 1 [O]$_{145}$ & 5.8E+16 & 7.8 & 2.1 & 3.4E+17 \\
Comp. 2 [O]$_{145}$ & 1.8E+17 & 10.0 & 4.7 & 1.1E+18 \\
Comp. 3 [O]$_{145}$ & 2.4E+16 & 11.7 & 11.0 & 1.4E+17 \\
\multicolumn{4}{l}{T$_{\mathrm{ex}}$ Background [OI]$_{63}$ = 77.5 K} \\
Comp. 1 [O]$_{63}$\tnote{c} & 3.4E+17 & 7.8 & 1.7 & \textbf{1.1E+19} \\
Comp. 2 [O]$_{63}$ & 3.3E+13 & 9.5 & 0.2 & 1.0E+15 \\
Comp. 3 [O]$_{63}$\tnote{c} & 1.1E+18 & 10.1 & 3.8 & \textbf{3.3E+19} \\
Comp. 4 [O]$_{63}$\tnote{c} & 1.4E+17 & 11.8 & 9.6 & \textbf{4.4E+18} \\
Comp. 5 [O]$_{63}$ & 2.3E+15 & 15.9 & 2.4 & 7.2E+16 \\
Comp. 6 [O]$_{63}$ & 3.1E+15 & 20.7 & 4.1 & 9.6E+16 \\
Comp. 7 [O]$_{63}$ & 3.1E+15 & 23.9 & 7.3 & 9.6E+16 \\

\hline
\multicolumn{4}{l}{T$_{\mathrm{ex}}$ Foreground = 19.9 K} \\
Comp. 1 [O]$_{63}$ & 6.5E+12 & 2.2 & 2.9 & \textbf{1.0E+18} \\
Comp. 2 [O]$_{63}$ & 9.5E+12 & 8.6 & 2.4 & \textbf{1.5E+18} \\
Comp. 3 [O]$_{63}$\tnote{d} & 1.0E+06 & 8.8 & 3.8 & 1.6E+11 & 5.9E+17\\
Comp. 4 [O]$_{63}$\tnote{d} & 1.4E+13 & 11.4 & 1.4 & \textbf{2.2E+18} & 2.1E+17 \\
\hline
\end{tabular}
\begin{tablenotes}\footnotesize
\item[a] Column density of the lower level population of the transition, calculated from  N$_{\mathrm{u}}$\Oone~and T$_{\mathrm{ex}}$.
\item[b] \Cp{} foreground column densities from \citet{2020A&A...636A..16G} when the velocities match.
\item[c] Components derived from the \OI{145} Gaussian fitting.
\item[d] Parameters of the components taken from the \Cp{} multi-component analysis from \citet{2020A&A...636A..16G}.
\end{tablenotes}
\end{threeparttable}
\label{app:monr2_1}
\end{table*}

\begin{table*}[ht]
\centering
\caption{Mon~R2 position 2 parameters.}
 \begin{threeparttable}
\begin{tabular}{c c c c | c | c }
\hline
\hline
Component & N$_u$ \Oone & V & $\Delta$V & N$_l$ \Oone\tnote{a} & N \Cp{}\tnote{b} \\
 &  (cm$^{-2}$) & (km s$^{-1}$) & (km~s$^{-1}$¸) & (cm$^{-2}$) & (cm$^{-2})$  \\
\hline
\multicolumn{4}{l}{T$_{\mathrm{ex}}$ Background [OI]$_{145}$ = 227.0 K} \\
Comp. 1 [O]$_{145}$ & 2.0E+17 & 9.9 & 6.3 & 9.3E+17 \\
Comp. 2 [O]$_{145}$ & 8.9E+16 & 11.5 & 2.8 & 4.1E+17 \\
\multicolumn{4}{l}{T$_{\mathrm{ex}}$ Background [OI]$_{63}$ = 140.0 K} \\
Comp. 1 [O]$_{63}$ & 1.8E+15 & 1.6 & 2.0 & 1.5E+16 \\
Comp. 2 [O]$_{63}$\tnote{c} & 9.3E+17 & 10.0 & 5.1 & \textbf{7.9E+18} \\
Comp. 3 [O]$_{63}$\tnote{c} & 4.1E+17 & 11.6 & 3.4 & \textbf{3.5E+18} \\
Comp. 4 [O]$_{63}$ & 1.1E+16 & 15.7 & 2.1 & 9.0E+16 \\
Comp. 5 [O]$_{63}$ & 9.1E+15 & 17.5 & 3.5 & 7.7E+16 \\
Comp. 6 [O]$_{63}$ & 5.6E+15 & 20.7 & 5.0 & 4.8E+16 \\
Comp. 7 [O]$_{63}$ & 2.7E+15 & 25.3 & 6.3 & 2.3E+16 \\
\hline
\multicolumn{4}{l}{T$_{\mathrm{ex}}$ Foreground = 20.0 K} \\
Comp. 1 [O]$_{63}$\tnote{d} & 5.7E+12 & 7.3 & 1.8 & \textbf{8.9E+17} & 9E+16 \\
Comp. 2 [O]$_{63}$ & 7.0E+12 & 7.6 & 4.1 & \textbf{1.1E+18} \\
Comp. 3 [O]$_{63}$\tnote{d} & 2.5E+07 & 9.7 & 3.7 & 4.0E+12 & 1.4E+17 \\
Comp. 4 [O]$_{63}$ & 1.4E+12 & 9.9 & 1.2 & 2.2E+17 \\
Comp. 5 [O]$_{63}$\tnote{d} & 4.4E+12 & 11.1 & 1.1 & \textbf{6.8E+17} & 2.5E+17 \\
Comp. 6 [O]$_{63}$\tnote{d} & 6.5E+12 & 12.0 & 1.1 & \textbf{1.0E+18} & 1.6E+17\\
\hline
\end{tabular}
\begin{tablenotes}\footnotesize
\item[a] Column density of the lower level population of the transition, calculated from  N$_{\mathrm{u}}$\Oone~and T$_{\mathrm{ex}}$.
\item[b] \Cp{} foreground column densities from \citet{2020A&A...636A..16G} when the velocities match.
\item[c] Components derived from the \OI{145} Gaussian fitting.
\item[d] Parameters of the components taken from the \Cp{} multi-component analysis from \citet{2020A&A...636A..16G}.
\end{tablenotes}
\end{threeparttable}
\label{app:monr2_1a}
\end{table*}

\begin{table*}[ht]
\centering
\caption{M17~SW position 0 parameters.}
\begin{threeparttable}
\begin{tabular}{c c c c | c | c }
\hline
\hline
Component & N$_u$ \Oone & V & $\Delta$V & N$_l$ \Oone\tnote{a} & N \Cp{}\tnote{b} \\
 &  (cm$^{-2}$) & (km~s$^{-1}$¸) & (km~s$^{-1}$¸) & (cm$^{-2}$) & (cm$^{-2})$  \\
\hline
\multicolumn{4}{l}{T$_{\mathrm{ex}}$ Background [OI]$_{145}$ = 200.1 K} \\
Comp. 1 [O]$_{145}$ & 7.0E+16 & 20.3 & 3.2 & 3.4E+17 \\
Comp. 2 [O]$_{145}$ & 5.7E+16 & 20.7 & 5.2 & 2.8E+17 \\
Comp. 3 [O]$_{145}$ & 1.0E+16 & 21.6 & 2.5 & 5.2E+16 \\
\multicolumn{4}{l}{T$_{\mathrm{ex}}$ Background [OI]$_{63}$ = 94.9 K} \\
Comp. 1 [O]$_{63}$ & 3.6E+15 & 12.1 & 3.5 & 6.5E+16 \\
Comp. 2 [O]$_{63}$ & 2.6E+15 & 14.5 & 1.7 & 4.8E+16 \\
Comp. 3 [O]$_{63}$\tnote{c} & 3.4E+17 & 20.3 & 2.6 & \textbf{6.3E+18} \\
Comp. 4 [O]$_{63}$\tnote{c} & 2.8E+17 & 20.7 & 4.1 & \textbf{5.1E+18} \\
Comp. 5 [O]$_{63}$\tnote{c} & 5.1E+16 & 21.5 & 2.0 & 9.4E+17 \\
Comp. 6 [O]$_{63}$ & 1.3E+15 & 26.3 & 1.2 & 2.3E+16 \\
Comp. 7 [O]$_{63}$ & 1.0E+15 & 27.9 & 1.1 & 1.9E+16 \\
\hline
\multicolumn{4}{l}{T$_{\mathrm{ex}}$ Foreground = 21.0 K} \\
Comp. 1 [O]$_{63}$\tnote{d} & 3.4E+10 & 17.0 & 0.1 & 2.9E+15 & 6.5E+17 \\
Comp. 2 [O]$_{63}$ & 2.3E+12 & 17.4 & 2.1 & \textbf{1.9E+17} \\
Comp. 3 [O]$_{63}$\tnote{d} & 9.2E+06 & 19.0 & 3.5 & 7.8E+11 & 3.5E+17\\
Comp. 4 [O]$_{63}$\tnote{d} & 2.0E+07 & 21.4 & 2.3 & 1.7E+12 & 7.4E+17 \\
Comp. 5 [O]$_{63}$ & 6.0E+12 & 21.7 & 1.7 & \textbf{5.1E+17} \\
Comp. 6 [O]$_{63}$ & 8.2E+11 & 23.2 & 0.2 & 2.9E+16 \\
Comp. 7 [O]$_{63}$\tnote{d} & 1.4E+13 & 24.1 & 0.6 & \textbf{1.1E+18} & 3.2E+17 \\
\hline
\end{tabular}
\begin{tablenotes}\footnotesize
\item[a] Column density of the lower level population of the transition, calculated from  N$_{\mathrm{u}}$\Oone~and T$_{\mathrm{ex}}$.
\item[b] \Cp{} foreground column densities from \citet{2020A&A...636A..16G} when the velocities match.
\item[c] Components derived from the \OI{145} Gaussian fitting.
\item[d] Parameters of the components taken from the \Cp{} multi-component analysis from \citet{2020A&A...636A..16G}.
\end{tablenotes}
\end{threeparttable}
\label{app:m17_1}
\end{table*}

\begin{table*}[ht]
\centering
\caption{M17~SW position 1 parameters.}
\begin{threeparttable}
\begin{tabular}{c c c c | c | c }
\hline
\hline
Component & N$_u$ \Oone & V & $\Delta$V & N$_l$ \Oone\tnote{a} & N \Cp{}\tnote{b} \\
 &  (cm$^{-2}$) & (km~s$^{-1}$¸) & (km~s$^{-1}$¸) & (cm$^{-2}$) & (cm$^{-2})$  \\
\hline
\multicolumn{4}{l}{T$_{\mathrm{ex}}$ Background [OI]$_{145}$ = 112.6 K} \\
Comp. 1 [O]$_{145}$ & 1.2E+17 & 18.9 & 4.1 & 8.8E+17 \\
\multicolumn{4}{l}{T$_{\mathrm{ex}}$ Background [OI]$_{63}$ = 65.0 K} \\
Comp. 1 [O]$_{63}$ & 1.2E+16 & 12.5 & 3.8 & 6.8E+17 \\
Comp. 2 [O]$_{63}$\tnote{c} & 8.7E+17 & 18.8 & 3.3 & \textbf{4.8E+19} \\
\hline
\multicolumn{4}{l}{T$_{\mathrm{ex}}$ Foreground = 20.8 K} \\
Comp. 1 [O]$_{63}$\tnote{d} & 5.6E+12 & 16.4 & 3.3 & \textbf{5.3E+17} & 5.3E+17\\
Comp. 2 [O]$_{63}$\tnote{d} & 1.7E+13 & 18.7 & 3.5 & \textbf{1.6E+18} & 4.5E+17\\
Comp. 3 [O]$_{63}$\tnote{d} & 1.0E+13 & 21.3 & 2.3 & \textbf{9.8E+17} & 5.8E+17\\
Comp. 4 [O]$_{63}$\tnote{d} & 1.9E+12 & 24.1 & 0.5 & \textbf{1.6E+17} & 9.1E+16 \\
\hline
\end{tabular}
\begin{tablenotes}\footnotesize
\item[a] Column density of the lower level population of the transition, calculated from  N$_{\mathrm{u}}$\Oone~and T$_{\mathrm{ex}}$.
\item[b] \Cp{} foreground column densities from \citet{2020A&A...636A..16G} when the velocities match.
\item[c] Components derived from the \OI{145} Gaussian fitting.
\item[d] Parameters of the components taken from the \Cp{} multi-component analysis from \citet{2020A&A...636A..16G}.
\end{tablenotes}
\end{threeparttable}
\label{app:m17_2}
\end{table*}

\begin{table*}[ht]
\centering
\caption{M17~SW position 2 parameters.}
\begin{threeparttable}
\begin{tabular}{c c c c | c | c }
\hline
\hline
Component & N$_u$ \Oone & V & $\Delta$V & N$_l$ \Oone\tnote{a} & N \Cp{}\tnote{b} \\
 &  (cm$^{-2}$) & (km~s$^{-1}$¸) & (km~s$^{-1}$¸) & (cm$^{-2}$) & (cm$^{-2})$  \\
\hline
\multicolumn{4}{l}{T$_{\mathrm{ex}}$ Background [OI]$_{145}$ = 125.1 K} \\
Comp. 1 [O]$_{145}$ & 1.3E+17 & 19.7 & 4.7 & 8.8E+17 \\
\multicolumn{4}{l}{T$_{\mathrm{ex}}$ Background [OI]$_{63}$ = 70.0 K} \\
Comp. 1 [O]$_{63}$ & 7.2E+15 & 12.6 & 3.9 & 3.1E+17 \\
Comp. 2 [O]$_{63}$\tnote{c} & 8.8E+17 & 19.6 & 3.8 & \textbf{3.8E+19} \\
\hline
\multicolumn{4}{l}{T$_{\mathrm{ex}}$ Foreground = 19.0 K} \\
Comp. 1 [O]$_{63}$\tnote{d} & 1.0E+07 & 14.0 & 3.0 & 2.7E+12 & 5.5E+17\\
Comp. 2 [O]$_{63}$ & 1.9E+12 & 16.0 & 3.2 & \textbf{5.0E+12} & 9.1E+17 \\
Comp. 3 [O]$_{63}$\tnote{d} & 5.6E+10 & 16.7 & 0.5 & 1.5E+16 \\
Comp. 4 [O]$_{63}$\tnote{d} & 3.0E+12 & 19.1 & 3.1 & \textbf{8.1E+17} & 3.5E+17 \\
Comp. 5 [O]$_{63}$\tnote{d} & 2.5E+12 & 21.2 & 2.0 & \textbf{6.7E+17} & 5.5E+17\\
Comp. 6 [O]$_{63}$\tnote{d} & 5.2E+12 & 24.3 & 2.8 & \textbf{1.4E+18} & 6.1E+17 \\
\hline
\end{tabular}
\begin{tablenotes}\footnotesize
\item[a] Column density of the lower level population of the transition, calculated from  N$_{\mathrm{u}}$\Oone~and T$_{\mathrm{ex}}$.
\item[b] \Cp{} foreground column densities from \citet{2020A&A...636A..16G} when the velocities match.
\item[c] Components derived from the \OI{145} Gaussian fitting.
\item[d] Parameters of the components taken from the \Cp{} multi-component analysis from \citet{2020A&A...636A..16G}.
\end{tablenotes}
\end{threeparttable}
\label{app:m17_3}
\end{table*}

\begin{table*}[ht]
\centering
\caption{M17~SW position 3 parameters.}
\begin{threeparttable}
\begin{tabular}{c c c c | c | c }
\hline
\hline
Component & N$_u$ \Oone & V & $\Delta$V & N$_l$ \Oone\tnote{a} & N \Cp{}\tnote{b} \\
 &  (cm$^{-2}$) & (km~s$^{-1}$¸) & (km~s$^{-1}$¸) & (cm$^{-2}$) & (cm$^{-2})$  \\
\hline
\multicolumn{4}{l}{T$_{\mathrm{ex}}$ Background [OI]$_{145}$ = 170.4 K} \\
Comp. 1 [O]$_{145}$ & 6.1E+16 & 20.0 & 4.6 & 3.3E+17 \\
Comp. 2 [O]$_{145}$ & 3.9E+16 & 23.7 & 3.6 & 2.1E+17 \\
\multicolumn{4}{l}{T$_{\mathrm{ex}}$ Background [OI]$_{63}$ = 86.0 K} \\
Comp. 1 [O]$_{63}$ & 2.2E+15 & 9.5 & 3.7 & 5.3E+16 \\
Comp. 2 [O]$_{63}$ & 3.2E+17 & 20.0 & 3.7 & \textbf{7.6E+18} \\
Comp. 3 [O]$_{63}$ & 2.1E+17 & 23.7 & 2.9 &\textbf{4.9E+18} \\
\hline
\multicolumn{4}{l}{T$_{\mathrm{ex}}$ Foreground = 19.6 K} \\
Comp. 1 [O]$_{63}$ & 9.4E+11 & 16.9 & 1.6 & 1.5E+17 \\
Comp. 2 [O]$_{63}$ & 1.0E+12 & 21.3 & 1.1 & 1.5E+17 \\
Comp. 3 [O]$_{63}$\tnote{d} & 5.6E+12 & 21.3 & 2.0 & \textbf{8.9E+17} & 5.1E+17\\
Comp. 4 [O]$_{63}$\tnote{d} & 7.7E+12 & 23.7 & 1.5 & \textbf{1.2E+18} & 2.6E+17\\
Comp. 5 [O]$_{63}$ & 3.6E+12 & 25.9 & 2.2 & 5.7E+17 \\
\hline
\end{tabular}
\begin{tablenotes}\footnotesize
\item[a] Column density of the lower level population of the transition, calculated from  N$_{\mathrm{u}}$\Oone~and T$_{\mathrm{ex}}$.
\item[b] \Cp{} foreground column densities from \citet{2020A&A...636A..16G} when the velocities match.
\item[c] Components derived from the \OI{145} Gaussian fitting.
\item[d] Parameters of the components taken from the \Cp{} multi-component analysis from \citet{2020A&A...636A..16G}.
\end{tablenotes}
\end{threeparttable}
\label{app:m17_4}
\end{table*}

\begin{table*}[ht]
\centering
\caption{M17~SW position 4 parameters.}
\begin{threeparttable}
\begin{tabular}{c c c c | c | c }
\hline
\hline
Component & N$_u$ \Oone & V & $\Delta$V & N$_l$ \Oone\tnote{a} & N \Cp{}\tnote{b} \\
 &  (cm$^{-2}$) & (km~s$^{-1}$¸) & (km~s$^{-1}$¸) & (cm$^{-2}$) & (cm$^{-2})$  \\
\hline
\multicolumn{4}{l}{T$_{\mathrm{ex}}$ Background [OI]$_{145}$ = 238.6 K} \\
Comp. 1 [O]$_{145}$ & 4.5E+16 & 20.0 & 2.5 & 2.0E+17 \\
Comp. 2 [O]$_{145}$ & 1.9E+17 & 21.3 & 6.2 & 8.4E+17 \\
\multicolumn{4}{l}{T$_{\mathrm{ex}}$ Background [OI]$_{63}$ = 145.0 K} \\
Comp. 1 [O]$_{63}$ & 2.3E+16 & 16.5 & 4.0 & 1.8E+17 \\
Comp. 2 [O]$_{63}$\tnote{c} & 2.0E+17 & 20.0 & 3.1 & \textbf{1.6E+18} \\
Comp. 3 [O]$_{63}$\tnote{c} & 8.4E+17 & 21.3 & 5.0 & \textbf{6.7E+18} \\
\hline
\multicolumn{4}{l}{T$_{\mathrm{ex}}$ Foreground = 20.7 K} \\
Comp. 1 [O]$_{63}$\tnote{d} & 1.3E+07 & 16.6 & 3.0 & 1.3E+12 & 2.7E+17\\
Comp. 2 [O]$_{63}$ & 4.6E+07 & 19.6 & 2.1 & 4.6E+12 \\
Comp. 3 [O]$_{63}$\tnote{d} & 1.3E+12 & 19.6 & 1.2 & 1.2E+17 & 2.1E+17\\
Comp. 4 [O]$_{63}$\tnote{d} & 4.6E+12 & 21.4 & 2.4 & \textbf{4.6E+17} & 3.8E+17 \\
Comp. 5 [O]$_{63}$ & 9.9E+12 & 21.5 & 0.8 & 9.8E+16 \\
Comp. 6 [O]$_{63}$\tnote{d} & 6.7E+12 & 23.8 & 1.1 & \textbf{6.5E+17} & 1.4E+17 \\
Comp. 7 [O]$_{63}$ & 1.0E+13 & 24.6 & 4.1 & \textbf{1.0E+18} \\
Comp. 8 [O]$_{63}$\tnote{d} & 1.4E+12 & 26.8 & 3.4 & 1.4E+17 & 2.6E+17\\
\hline
\end{tabular}
\begin{tablenotes}\footnotesize
\item[a] Column density of the lower level population of the transition, calculated from  N$_{\mathrm{u}}$\Oone~and T$_{\mathrm{ex}}$.
\item[b] \Cp{} foreground column densities from \citet{2020A&A...636A..16G} when the velocities match.
\item[c] Components derived from the \OI{145} Gaussian fitting.
\item[d] Parameters of the components taken from the \Cp{} multi-component analysis from \citet{2020A&A...636A..16G}.
\end{tablenotes}
\end{threeparttable}
\label{app:m17_5}
\end{table*}

\begin{table*}[ht]
\centering
\caption{M17~SW position 5 parameters.}
\begin{threeparttable}
\begin{tabular}{c c c c | c | c }
\hline
\hline
Component & N$_u$ \Oone & V & $\Delta$V & N$_l$ \Oone\tnote{a} & N \Cp{}\tnote{b} \\
 &  (cm$^{-2}$) & (km~s$^{-1}$¸) & (km~s$^{-1}$¸) & (cm$^{-2}$) & (cm$^{-2})$  \\
\hline
\multicolumn{4}{l}{T$_{\mathrm{ex}}$ Background [OI]$_{145}$ = 117.3 K} \\
Comp. 1 [O]$_{145}$ & 1.3E+16 & 13.6 & 4.0 & 8.7E+16 \\
Comp. 2 [O]$_{145}$ & 6.3E+15 & 21.1 & 2.3 & 4.4E+16 \\
Comp. 3 [O]$_{145}$ & 1.7E+16 & 22.9 & 7.3 & 1.2E+17 \\
\multicolumn{4}{l}{T$_{\mathrm{ex}}$ Background [OI]$_{63}$ = 66.9 K} \\
Comp. 1 [O]$_{63}$\tnote{c} & 8.7E+16 & 13.6 & 3.2 & \textbf{4.3E+18} \\
Comp. 2 [O]$_{63}$\tnote{c} & 4.3E+16 & 21.2 & 1.9 & \textbf{2.2E+18} \\
Comp. 3 [O]$_{63}$\tnote{c} & 1.2E+17 & 22.8 & 5.8 & \textbf{5.9E+18} \\
\hline
\multicolumn{4}{l}{T$_{\mathrm{ex}}$ Foreground = 20.0 K} \\
Comp. 1 [O]$_{63}$\tnote{d} & 1.1E+13 & 12.7 & 2.7 & \textbf{1.6E+18} & 1.8E+17 \\
Comp. 2 [O]$_{63}$\tnote{d} & 1.2E+12 & 18.4 & 2.0 & 1.8E+17 & 1.4E+17 \\
Comp. 3 [O]$_{63}$ & 3.7E+12 & 19.9 & 2.7 & \textbf{5.5E+17} \\
Comp. 4 [O]$_{63}$ & 9.0E+12 & 23.8 & 6.0 & \textbf{1.4E+18} \\
Comp. 5 [O]$_{63}$\tnote{d} & 1.6E+12 & 23.9 & 0.8 & 2.4E+17 & 7.2E+16 \\
\hline
\end{tabular}
\begin{tablenotes}\footnotesize
\item[a] Column density of the lower level population of the transition, calculated from  N$_{\mathrm{u}}$\Oone~and T$_{\mathrm{ex}}$.
\item[b] \Cp{} foreground column densities from \citet{2020A&A...636A..16G} when the velocities match.
\item[c] Components derived from the \OI{145} Gaussian fitting.
\item[d] Parameters of the components taken from the \Cp{} multi-component analysis from \citet{2020A&A...636A..16G}.
\end{tablenotes}
\end{threeparttable}
\label{app:m17_6}
\end{table*}

\begin{table*}[ht]
\centering
\caption{M17~SW position 6 parameters.}
\begin{threeparttable}
\begin{tabular}{c c c c | c | c }
\hline
\hline
Component & N$_u$ \Oone & V & $\Delta$V & N$_l$ \Oone\tnote{a} & N \Cp{}\tnote{b} \\
 &  (cm$^{-2}$) & (km~s$^{-1}$¸) & (km~s$^{-1}$¸) & (cm$^{-2}$) & (cm$^{-2})$  \\
\hline
\multicolumn{4}{l}{T$_{\mathrm{ex}}$ Background [OI]$_{145}$ = 221.6 K} \\
Comp. 1 [O]$_{145}$ & 9.3E+16 & 19.8 & 7.2 & 4.4E+17 \\
Comp. 2 [O]$_{145}$ & 9.4E+16 & 21.5 & 2.9 & 4.4E+17 \\
\multicolumn{4}{l}{T$_{\mathrm{ex}}$ Background [OI]$_{63}$ = 101.0 K} \\
Comp. 1 [O]$_{63}$\tnote{b} & 4.4E+17 & 19.8 & 5.8 & \textbf{6.9E+18} \\
Comp. 2 [O]$_{63}$\tnote{b} & 4.4E+17 & 21.5 & 2.3 & \textbf{7.0E+18} \\
\hline
\multicolumn{4}{l}{T$_{\mathrm{ex}}$ Foreground = 20.2 K} \\
Comp. 1 [O]$_{63}$ & 1.0E+13 & 15.5 & 4.0 & \textbf{1.4E+18} \\
Comp. 2 [O]$_{63}$\tnote{c} & 3.5E+12 & 17.2 & 3.1 & \textbf{4.7E+17} & 5.7E+17\\
Comp. 3 [O]$_{63}$\tnote{c} & 1.0E+12 & 19.1 & 2.2 & 1.4E+17 & 3.2E+17 \\
Comp. 4 [O]$_{63}$\tnote{c} & 2.2E+12 & 21.6 & 1.6 & 2.9E+17 & 7.6E+17\\
Comp. 5 [O]$_{63}$ & 3.0E+08 & 21.6 & 2.3 & 3.9E+13 \\
Comp. 6 [O]$_{63}$\tnote{c} & 2.0E+12 & 24.1 & 0.7 & 2.5E+17 & 1.4E+17 \\
\hline
\end{tabular}
\begin{tablenotes}\footnotesize
\item[a] Column density of the lower level population of the transition, calculated from  N$_{\mathrm{u}}$\Oone~and T$_{\mathrm{ex}}$.
\item[b] \Cp{} foreground column densities from \citet{2020A&A...636A..16G} when the velocities match.
\item[c] Components derived from the \OI{145} Gaussian fitting.
\item[d] Parameters of the components taken from the \Cp{} multi-component analysis from \citet{2020A&A...636A..16G}.
\end{tablenotes}
\end{threeparttable}
\label{app:m17_7}
\end{table*}

\begin{table*}[ht]
\centering
\caption{Mon~R2 position 1 parameters using free foreground parameters of position 2.}
\begin{threeparttable}
\begin{tabular}{c c c c | c }
\hline
\hline
Component & N$_u$ \Oone & V & $\Delta$V & N$_l$ \Oone\tnote{a}  \\
 &  (cm$^{-2}$) & (km~s$^{-1}$¸) & (km~s$^{-1}$¸) & (cm$^{-2}$)  \\
\hline
\multicolumn{4}{l}{T$_{\mathrm{ex}}$ Background [OI]$_{145}$ = 144.6 K} \\
Comp. 1 [O]$_{145}$ & 5.8E+16 & 7.8 & 2.1 & 3.4E+17 \\
Comp. 2 [O]$_{145}$ & 1.8E+17 & 10.0 & 4.7 & 1.1E+18 \\
Comp. 3 [O]$_{145}$ & 2.4E+16 & 11.7 & 11.0 & 1.4E+17 \\
\multicolumn{4}{l}{T$_{\mathrm{ex}}$ Background [OI]$_{63}$ = 77.5 K} \\
Comp. 1 [O]$_{63}$\tnote{b} & 3.4E+17 & 7.8 & 1.7 & \textbf{1.1E+19} \\
Comp. 2 [O]$_{63}$ & 1.6E+16 & 9.4 & 1.0 & 5.0E+17 \\
Comp. 3 [O]$_{63}$\tnote{b} & 1.1E+18 & 10.1 & 3.8 & \textbf{3.3E+19} \\
Comp. 4 [O]$_{63}$\tnote{b} & 1.4E+17 & 11.8 & 9.6 & 4.4E+18 \\
Comp. 5 [O]$_{63}$ & 2.3E+15 & 15.9 & 2.4 & 7.3E+16 \\
Comp. 6 [O]$_{63}$ & 3.1E+15 & 20.7 & 4.1 & 9.7E+16 \\
Comp. 7 [O]$_{63}$ & 3.0E+15 & 23.9 & 7.3 & 9.5E+16 \\
\hline
\multicolumn{4}{l}{T$_{\mathrm{ex}}$ Foreground = 19.9 K} \\
Comp. 1 [O]$_{63}$ & 5.7E+12 & 7.3 & 1.8 & \textbf{8.9E+17} \\
Comp. 2 [O]$_{63}$ & 7.0E+12 & 7.6 & 4.1 & \textbf{1.1E+18} \\
Comp. 3 [O]$_{63}$ & 2.5E+07 & 9.7 & 3.7 & 4.0E+12 \\
Comp. 4 [O]$_{63}$ & 1.4E+12 & 9.9 & 1.2 & 2.2E+17 \\
Comp. 5 [O]$_{63}$ & 4.4E+12 & 11.1 & 1.1 & \textbf{6.8E+17} \\
Comp. 6 [O]$_{63}$ & 6.5E+12 & 12.0 & 1.1 & \textbf{1.0E+18} \\
\hline 
\end{tabular}
\begin{tablenotes}\footnotesize
\item[a] Column density of the lower level population of the transition, calculated from  N$_{\mathrm{u}}$\Oone~and T$_{\mathrm{ex}}$.
\item[b] Components derived from the \OI{145} Gaussian fitting.
\end{tablenotes}
\end{threeparttable}
\label{app:monr2_1_fixed1}
\end{table*}

\begin{table*}[ht]
\centering
\caption{Mon~R2 position 1 parameters using free foreground parameters of position 2.}
\begin{threeparttable}
\begin{tabular}{c c c c | c }
\hline
\hline
Component & N$_u$ \Oone & V & $\Delta$V & N$_l$ \Oone\tnote{a} \\
 &  (cm$^{-2}$) & (km~s$^{-1}$¸) & (km~s$^{-1}$¸) & (cm$^{-2}$)  \\
\hline
\multicolumn{4}{l}{T$_{\mathrm{ex}}$ Background [OI]$_{145}$ = 144.6 K} \\
Comp. 1 [O]$_{145}$ & 5.8E+16 & 7.8 & 2.1 & 3.4E+17 \\
Comp. 2 [O]$_{145}$ & 1.8E+17 & 10.0 & 4.7 & 1.1E+18 \\
Comp. 3 [O]$_{145}$ & 2.4E+16 & 11.7 & 11.0 & 1.4E+17 \\
\multicolumn{4}{l}{T$_{\mathrm{ex}}$ Background [OI]$_{63}$ = 77.5 K} \\
Comp. 1 [O]$_{63}$\tnote{b} & 3.4E+17 & 7.8 & 1.7 & \textbf{1.1E+19} \\
Comp. 2 [O]$_{63}$ & 1.6E+16 & 9.4 & 1.0 & 5.0E+17 \\
Comp. 3 [O]$_{63}$\tnote{b} & 1.1E+18 & 10.1 & 3.8 & \textbf{3.3E+19} \\
Comp. 4 [O]$_{63}$\tnote{b} & 1.4E+17 & 11.8 & 9.6 & 4.4E+18 \\
Comp. 5 [O]$_{63}$ & 2.3E+15 & 15.9 & 2.4 & 7.3E+16 \\
Comp. 6 [O]$_{63}$ & 3.1E+15 & 20.7 & 4.1 & 9.7E+16 \\
Comp. 7 [O]$_{63}$ & 3.0E+15 & 23.9 & 7.3 & 9.5E+16 \\
\hline
\multicolumn{4}{l}{T$_{\mathrm{ex}}$ Foreground = 20.4 K} \\
Comp. 1 [O]$_{63}$ & 2.6E+12 & 7.3 & 1.4 & \textbf{3.0E+17} \\
Comp. 2 [O]$_{63}$ & 9.3E+11 & 7.6 & 9.2 & 1.1E+17 \\
Comp. 3 [O]$_{63}$ & 2.5E+08 & 9.7 & 2.9 & 2.9E+13 \\
Comp. 4 [O]$_{63}$ & 1.4E+13 & 9.9 & 2.9 & \textbf{1.6E+18} \\
Comp. 5 [O]$_{63}$ & 7.1E+12 & 11.1 & 0.9 & \textbf{8.2E+17} \\
Comp. 6 [O]$_{63}$ & 3.8E+12 & 12.0 & 0.9 & \textbf{4.3E+17} \\
\hline
\end{tabular}
\begin{tablenotes}\footnotesize
\item[a] Column density of the lower level population of the transition, calculated from  N$_{\mathrm{u}}$\Oone~and T$_{\mathrm{ex}}$.
\item[b] Components derived from the \OI{145} Gaussian fitting.
\end{tablenotes}
\end{threeparttable}
\label{app:Monr2_1_fixed2} 
\end{table*}

\begin{table*}[ht]
\centering
\caption{M17~SW position 6 parameters using fixed foreground parameters of position 0.}
\begin{threeparttable}
\begin{tabular}{c c c c | c }
\hline
\hline
Component & N$_u$ \Oone & V & $\Delta$V & N$_l$ \Oone\tnote{a}  \\
 &  (cm$^{-2}$) & (km~s$^{-1}$¸) & (km~s$^{-1}$¸) & (cm$^{-2}$)  \\
\hline
\multicolumn{4}{l}{T$_{\mathrm{ex}}$ Background [OI]$_{145}$ = 221.6 K} \\
Comp. 1 [O]$_{145}$ & 9.3E+16 & 19.8 & 7.2 & 4.4E+17 \\
Comp. 2 [O]$_{145}$ & 9.4E+16 & 21.5 & 2.9 & 4.4E+17 \\
\multicolumn{4}{l}{T$_{\mathrm{ex}}$ Background [OI]$_{63}$ = 101.0 K} \\
Comp. 1 [O]$_{63}$ & 4.4E+17 & 19.8 & 5.8 & \textbf{6.9E+18} \\
Comp. 2 [O]$_{63}$ & 4.4E+17 & 21.5 & 2.3 & \textbf{7.0E+18} \\
\hline
\multicolumn{4}{l}{T$_{\mathrm{ex}}$ Foreground = 21.0 K} \\
Comp. 1 [O]$_{63}$ & 5.5E+11 & 16.9 & 0.7 & 4.6E+16 \\
Comp. 2 [O]$_{63}$ & 1.6E+12 & 17.8 & 1.7 & 1.3E+17 \\
Comp. 3 [O]$_{63}$ & 4.6E+07 & 19.0 & 3.5 & 4.0E+12 \\
Comp. 4 [O]$_{63}$ & 8.5E+07 & 21.4 & 2.3 & 7.2E+12 \\
Comp. 5 [O]$_{63}$ & 6.0E+12 & 21.7 & 1.7 & \textbf{5.1E+17} \\
Comp. 6 [O]$_{63}$ & 8.7E+11 & 23.2 & 0.2 & 2.8E+16 \\
Comp. 7 [O]$_{63}$ & 1.4E+13 & 24.1 & 0.6 & \textbf{1.1E+18} \\
\hline
\end{tabular}
\begin{tablenotes}\footnotesize
\item[a] Column density of the lower level population of the transition, calculated from  N$_{\mathrm{u}}$\Oone~and T$_{\mathrm{ex}}$.
\item[b] Components derived from the \OI{145} Gaussian fitting.
\end{tablenotes}
\end{threeparttable}
\label{app:m17_6_fixed1}
\end{table*}

\begin{table*}[ht]
\centering
\caption{M17~SW position 6 parameters using free foreground parameters of position 0.}
\begin{threeparttable}
\begin{tabular}{c c c c | c }
\hline
\hline
Component & N$_u$ \Oone & V & $\Delta$V & N$_l$ \Oone\tnote{a} \\
 &  (cm$^{-2}$) & (km~s$^{-1}$¸) & (km~s$^{-1}$¸) & (cm$^{-2}$)  \\
\hline
\multicolumn{4}{l}{T$_{\mathrm{ex}}$ Background [OI]$_{145}$ = 221.6 K} \\
Comp. 1 [O]$_{145}$ & 9.3E+16 & 19.8 & 7.2 & 4.4E+17 \\
Comp. 2 [O]$_{145}$ & 9.4E+16 & 21.5 & 2.9 & 4.4E+17 \\
\multicolumn{4}{l}{T$_{\mathrm{ex}}$ Background [OI]$_{63}$ = 101.0 K} \\
Comp. 1 [O]$_{63}$ & 4.4E+17 & 19.8 & 5.8 & \textbf{6.9E+18} \\
Comp. 2 [O]$_{63}$ & 4.4E+17 & 21.5 & 2.3 & \textbf{7.0E+18} \\
\hline
\multicolumn{4}{l}{T$_{\mathrm{ex}}$ Foreground = 20.0 K} \\
Comp. 1 [O]$_{63}$ & 5.5E+12 & 16.9 & 3.1 & \textbf{8.1E+17} \\
Comp. 2 [O]$_{63}$ & 5.0E+12 & 17.8 & 6.8 & \textbf{7.3E+17} \\
Comp. 3 [O]$_{63}$ & 4.6E+06 & 19.0 & 35.0 & 6.8E+11 \\
Comp. 4 [O]$_{63}$ & 8.5E+06 & 21.4 & 22.5 & 1.2E+12 \\
Comp. 5 [O]$_{63}$ & 1.3E+12 & 21.7 & 1.3 & \textbf{1.8E+17} \\
Comp. 6 [O]$_{63}$ & 8.7E+10 & 23.2 & 0.2 & 2.7E+15 \\
Comp. 7 [O]$_{63}$ & 1.7E+12 & 24.1 & 0.7 & \textbf{2.4E+17} \\
\hline
\end{tabular}
\begin{tablenotes}\footnotesize
\item[a] Column density of the lower level population of the transition, calculated from  N$_{\mathrm{u}}$\Oone~and T$_{\mathrm{ex}}$.
\item[b] Components derived from the \OI{145} Gaussian fitting.
\end{tablenotes}
\end{threeparttable}
\label{app:m17_6_fixed2}
\end{table*}

\end{appendix}

\end{document}